\documentclass[letter,twocolumn,noamsfonts,superscriptaddress,showpacs,aps,prl,nobibnotes,10pt,a4paper]{revtex4-1}
\usepackage[utf8]{inputenc}
\usepackage[pdftex]{graphicx}
\usepackage{amsmath}
\usepackage{amssymb}
\usepackage{placeins}
\usepackage{float}
\usepackage[paperwidth=208.026mm,paperheight=279.4mm,left=1.5cm,right=1.37cm,top=2cm,bottom=1.72cm]{geometry}
\usepackage{xfrac}

\usepackage[colorlinks=true]{hyperref}
\hypersetup{
  urlcolor     = blue,    
  linkcolor    = red,    
  citecolor    = blue      
}
\usepackage{ragged2e}
\newcommand{\etal}{\textit{et al.}}
\newcommand{\eca}{EuCd$_2$As$_2$}
\begin{document}
~\vspace{0.4cm}
\title{Coupling of magnetic order and charge transport in the candidate Dirac semimetal EuCd$_2$As$_2$}
\author{M. C. Rahn}
\thanks{Present affiliation: Los Alamos National Laboratory, Los~Alamos, New Mexico 87545, USA}
\email[]{rahn@lanl.gov}
\affiliation{Department of Physics, University of Oxford, Clarendon Laboratory, Oxford, OX1 3PU, United Kingdom}
\author{J.-R. Soh}
\affiliation{Department of Physics, University of Oxford, Clarendon Laboratory, Oxford, OX1 3PU, United Kingdom}
\author{S. Francoual}
\affiliation{Deutsches Elektronen-Synchrotron (DESY), Notkestra\ss e 85, 22607 Hamburg, Germany}
\author{L. S. I. Veiga}
\thanks{Present affiliation: London Centre for Nanotechnology and Department of Physics and Astronomy,
University College London, London WC1E 6BT, United Kingdom}
\affiliation{Deutsches Elektronen-Synchrotron (DESY), Notkestra\ss e 85, 22607 Hamburg, Germany}
\author{J. Strempfer}
\thanks{Present affiliation: Advanced Photon Source, Argonne National Laboratory, 9700 S. Cass Avenue, Argonne, IL 60439, USA}
\affiliation{Deutsches Elektronen-Synchrotron (DESY), Notkestra\ss e 85, 22607 Hamburg, Germany}
\author{J.~Mardegan}
\thanks{Present affiliation: Paul Scherrer Institut, 5232 Villigen, Switzerland}
\affiliation{Deutsches Elektronen-Synchrotron (DESY), Notkestra\ss e 85, 22607 Hamburg, Germany}
\author{D. Y. Yan}
\affiliation{Beijing National Laboratory for Condensed Matter Physics, Institute of Physics, Chinese Academy of Sciences, Beijing 100190, China}
\author{Y. F. Guo}
\affiliation{School of Physical Science and Technology, ShanghaiTech University, Shanghai 201210, China}
\affiliation{CAS Center for Excellence in Superconducting Electronics (CENSE), Shanghai 200050, China}
\author{Y. G. Shi}
\affiliation{Beijing National Laboratory for Condensed Matter Physics, Institute of Physics, Chinese Academy of Sciences, Beijing 100190, China}
\author{A. T. Boothroyd}
\email[]{a.boothroyd@physics.ox.ac.uk}
\affiliation{Department of Physics, University of Oxford, Clarendon Laboratory, Oxford, OX1 3PU, United Kingdom}

\date{\today}

\begin{abstract}
We use resonant elastic x-ray scattering to determine the evolution of magnetic order in EuCd$_2$As$_2$ below $T_\textrm{N}=9.5$\,K, as a function of temperature and applied magnetic field. We find an A-type antiferromagneticstructure with in-plane magnetic moments, and observe dramatic magnetoresistive effects associated with field-induced changes in the magnetic structure and domain populations. Our \textit{ab initio} electronic structure calculations indicate that the Dirac dispersion found in the nonmagnetic Dirac semimetal Cd$_3$As$_2$ is also present in EuCd$_2$As$_2$, but is gapped for $T < T_\textrm{N}$ due to the breaking of $C_3$ symmetry by the magnetic structure.
\end{abstract}


\maketitle

\section{I. Introduction}
The layered intermetallic \eca~is of interest owing to its structural relation to Cd$_3$As$_2$, the first identified example of a bulk Dirac semimetal~\cite{Liu2014b,Neupane2014}. It features similar networks of edge-sharing CdAs$_4$ tetrahedra, but in a layered rather than three-dimensional geometry and with interleaved planes of magnetic Eu$^{2+}$ ions. While ever more non-magnetic topological phases are being proposed and discovered, the search for materials that combine non-trivial electronic topology with strong correlations has advanced at a slower pace~\cite{Bansil2016,Armitage2018}. Magnetism may provide a handle with which to control the unusual transport properties of Dirac fermions, such as giant negative magnetoresistance~\cite{Li2016,Wang2013}. This could be a first step towards an application in electronic devices~\cite{Simejkal2017}.

\eca~was first synthesized (in polycrystalline form) in the search for novel pnictide Zintl phases by Artmann \etal~\cite{Artmann1996}. Pnictides with 122 stoichiometry and trigonal CaAl$_2$Si$_2$-type structure (space group $P\bar{3}m1$) are less common than related tetragonal materials, such as the superconductor parent-compound EuFe$_2$As$_2$\cite{Artmann1996,Ren2008,Paramanik2014}. A number of such trigonal compounds, related to \eca~by ionic substitutions on any site, have recently attracted interest due to their (high temperature) thermoelectric properties \cite{Guo2013,Min2015}. As in other Eu-based materials of this family \cite{Jiang2006,Goryunov2012,Zhang2010}, low temperature magnetometry of \eca~powders revealed ferromagnetic correlations (a positive Curie--Weiss temperature $\Theta_\text{CW}=9.3\,$K) and a large effective magnetic moment of $\mu_\textrm{eff} = 7.7\,\mu_\text{B}$, close to the free-ion value of $7.94\,\mu_\text{B}$ expected for the divalent state of Eu ($S=7/2$, $L=0$)~[\onlinecite{Artmann1996}].

The unusual magnetic transition at low temperatures was initially interpreted as a successive ferromagnetic (16\,K) and then antiferromagnetic (9.5\,K) ordering of the Eu spins in the triangular layers. In a M\"{o}ssbauer study of single-crystalline \eca, a ferromagnetic phase was later ruled out and attributed to Eu$^{3+}$-containing impurities introduced by oxidation of the polycrystalline samples\cite{Schellenberg2011}.

A first investigation of \eca~single crystals was recently reported by Wang \etal, who observed a strong coupling between charge transport and magnetic degrees of freedom\cite{Wang2016a}. This is seen as a strong resistivity anomaly associated with the magnetic ordering transition, which can be suppressed by an applied magnetic field.

In order to better understand the interplay between charge transport and magnetism, we have performed an in-depth resonant elastic x-ray scattering (REXS) study of the magnetic order in \eca, as a function of temperature and applied magnetic field. We then used these results to investigate what effect magnetic order has on the electronic structure as calculated by density functional theory.

Our REXS study establishes that EuCd$_2$As$_2$ orders into an A-type antiferromagnetic structure below $T_\textrm{N}=9.5$\,K, with the Eu spins aligned parallel to the triangular layers. This magnetic structure breaks the $C_3$ symmetry of the paramagnetic phase and, according to our \textit{ab initio} electronic structure calculations, opens a gap at the Dirac points in the vicinity of the Fermi surface. Conversely, these results imply that other materials of the CaAl$_2$Si$_2$--type structure could host Dirac fermions that couple to a magnetically ordered state at zero field, if the three-fold rotational symmetry is conserved.

\section{II. Experimental and Computational details}
\begin{figure}
\includegraphics[width=0.95\columnwidth,trim= 0pt 0pt 0pt 0pt, clip]{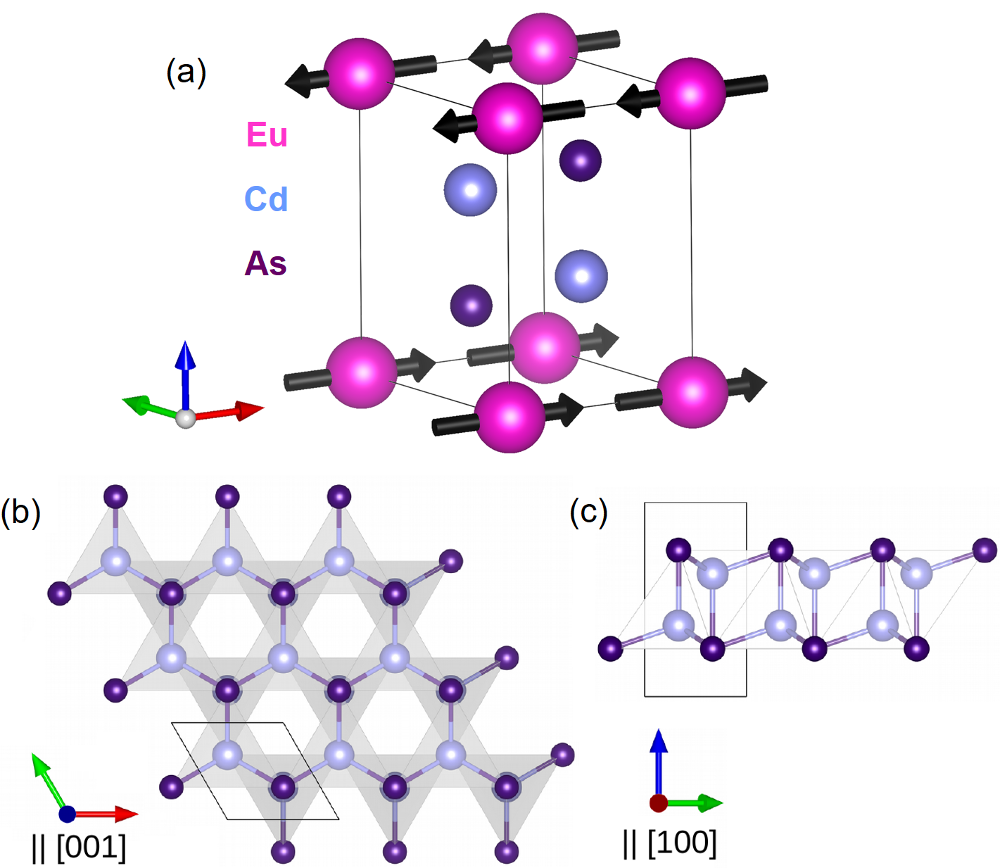}
\caption{\footnotesize \label{Fig1} (a) Structural unit cell of \eca~with $P\bar{3}m1$ trigonal symmetry and lattice parameters $a\simeq 4.44$\,\AA~and $c\simeq 7.33$\,\AA. The magnetic Eu ion is located at the origin (Wyckoff position \textbf{1a}), while Cd and As atoms at \textbf{2d} positions $(\sfrac13,\sfrac23, z_\textrm{Cd})$, $z_\textrm{Cd} \simeq 0.63$, and $(\sfrac13,\sfrac23, z_\textrm{As})$, $z_\textrm{As} \simeq 0.25$, form layers of Cd--As tetrahedra as illustrated in panels (b)--(c). The A-type antiferromagnetic structure, propagation vector $\mathbf{q}_\text{m}=(0,0,0.5)$, with in-plane magnetic moments is also shown.}
\end{figure}

Single crystals of \eca~were synthesized by the NaCl/KCl flux growth method, as described by Schellenberg \etal~\cite{Schellenberg2011}. This yields shiny metallic platelets ($c$ axis surface normal) with dimensions up to $2\times2\,\text{mm}^2$ and $0.5\,\text{mm}$ thickness. The samples were screened by Mo $K_\alpha$ x-ray diffraction~\cite{Agilent} to confirm the trigonal crystal structure (space group $P\bar{3}m1$) that had been originally determined by Artmann \etal~\cite{Artmann1996} (see Supplemental Material \cite{supplemental,Rodriguez1993}). The structure, along with the model of the zero-field magnetic order determined here, is illustrated in Fig.~\ref{Fig1}.

Measurements of the magnetic properties of \eca~had previously been reported for polycrystalline samples~\cite{Artmann1996}, and on single crystals with the field applied along an arbitrary direction~\cite{Schellenberg2011}. Here, we supplement these results with magnetometry data recorded in magnetic fields applied both parallel and perpendicular to the trigonal $c$ axis. These measurements were made with a superconducting quantum interference device (SQUID) magnetometer (Quantum Design~\cite{QuantumDesign}).

The thermal variation of the resistivity of \eca~single crystals has recently been reported by Wang \etal~\cite{Wang2016a} (at zero field and 9\,T). In order to characterize the electronic transport properties of \eca~in more detail, we performed alternating current transport (ACT) measurements using the six-point sample contacting geometry (Quantum Design PPMS; the experimental arrangement is described in the Supplemental Material~\cite{supplemental}). The magnetic field was applied along the $c$ axis throughout, and the current was applied in the $a$-$b$ plane.

In order to determine the magnetic structure of \eca, REXS experiments were performed in the second experimental hutch (EH2) of beamline P09 of PETRA-III at DESY; see Ref.~\onlinecite{Strempfer2013}. Neutron diffraction studies had been called for in the literature \cite{Wang2016a}, but are hindered by the extremely large neutron absorption cross-sections of both Eu (4530\,b) and Cd (2520\,b). On the other hand, the magnetic moments of rare earth ions are known to couple strongly to $L$ edge resonant dipole ($2p\,\rightarrow\,5d$) transitions. REXS in the hard x-ray regime, therefore, is an ideal alternative for magnetic structure determination in this material.

Two separate EuCd$_2$As$_2$ single crystals were probed, with a magnetic field applied perpendicular and nearly parallel to the $c$ axis, respectively. To enhance the magnetic x-ray scattering from Eu$^{2+}$ magnetic ions, the photon energy was tuned to the Eu $L_3$ edge at $\hbar\omega\approx6.972\,$ keV ($\lambda=1.778\,$\AA). The REXS setup at beamline P09-EH2 has several unique advantages for the present study, including a double phase retarder for full linear polarization analysis (FLPA) at hard x-ray energies and a vertical 14\,T cryomagnet. We provide detailed information on the sample orientation, scattering geometry~\cite{Detlefs2012,Busing1967} and FLPA technique~\cite{Mazzoli2007,Johnson2008,Hatton2009,Shukla2012,Francoual2015} in the Supplemental Material~\cite{supplemental}.

For full polarization analysis, two magnetic Bragg peaks were selected in each experimental setting. In the first setting ($\bf{H}\perp\bf{c}$), magnetic scattering at the wave vectors $(0,0,4.5)$ and $(1,0,5.5)$ (in reciprocal lattice units) was characterized in zero field as well as in applied fields of 50 and 300\,mT. In the second setting ($\bf{H}\parallel\bf{c}$) we investigated peaks at $(4,0,-0.5)$ and $(4,-2,-0.5)$, in zero field and in applied fields of 0.5 and 1.0\,T.

The REXS scattering amplitude has a complex dependence on the relative orientations of the magnetic moments, x-ray polarization and wave vectors. It can be conveniently expressed in a basis of polarization vectors perpendicular, $\bf{\sigma}$=(1,0), and parallel, $\bf{\pi}$=(0,1), to the scattering plane~\cite{Blume1988}. The definition of these vectors in the present scattering geometry is illustrated in Fig.~S3 of the Supplemental Material~\cite{supplemental}.

In the case of dominant electric dipole (E1) excitations, the scattering amplitude tensor is given by~\cite{Hill1996}
\begin{align}\label{Eq1}
\begin{split}
F_{E1}&\propto \left(\hat{\bf{\epsilon}}_f\times\hat{\bf{\epsilon}}_i\right)\cdot\hat{\bf{M}}_{\hat{\mathbf{u}}} = \begin{pmatrix} 0 & \hat{\bf{k}}_i\\ -\hat{\bf{k}}_f & \hat{\bf{k}}_f\times\hat{\bf{k}}_i \end{pmatrix} \cdot \hat{\bf{M}}_{\hat{\mathbf{u}}}\\ &= \begin{pmatrix} 0 & M_1^{\hat{\mathbf{u}}}\,\cos\theta +M_3^{\hat{\mathbf{u}}}\,\sin\theta \\ M_3^{\hat{\mathbf{u}}}\,\sin\theta-M_1^{\hat{\mathbf{u}}}\cos\theta & -M_2^{\hat{\mathbf{u}}}\,\sin 2\theta \end{pmatrix}~~,
\end{split}
\end{align}
where $2\theta$ is the scattering angle and $\hat{\bf{\epsilon}}_{i/f}$ and $\hat{\bf{k}}_{i/f}$ denote the directions of the incident and scattered linear polarization and wave vectors, respectively. $\hat{\bf{M}}_{\hat{\mathbf{u}}}$ refers to the magnetic structure factor vector in the conventional $\hat{\bf{u}}_i$ reference frame introduced by Blume and Gibbs~\cite{Blume1988} and defined in Eq.~S10 of the Supplemental Material~\cite{supplemental}.

\begin{figure*}
\includegraphics[width=2.0\columnwidth,trim= 6pt 6pt 92pt 25pt, clip]{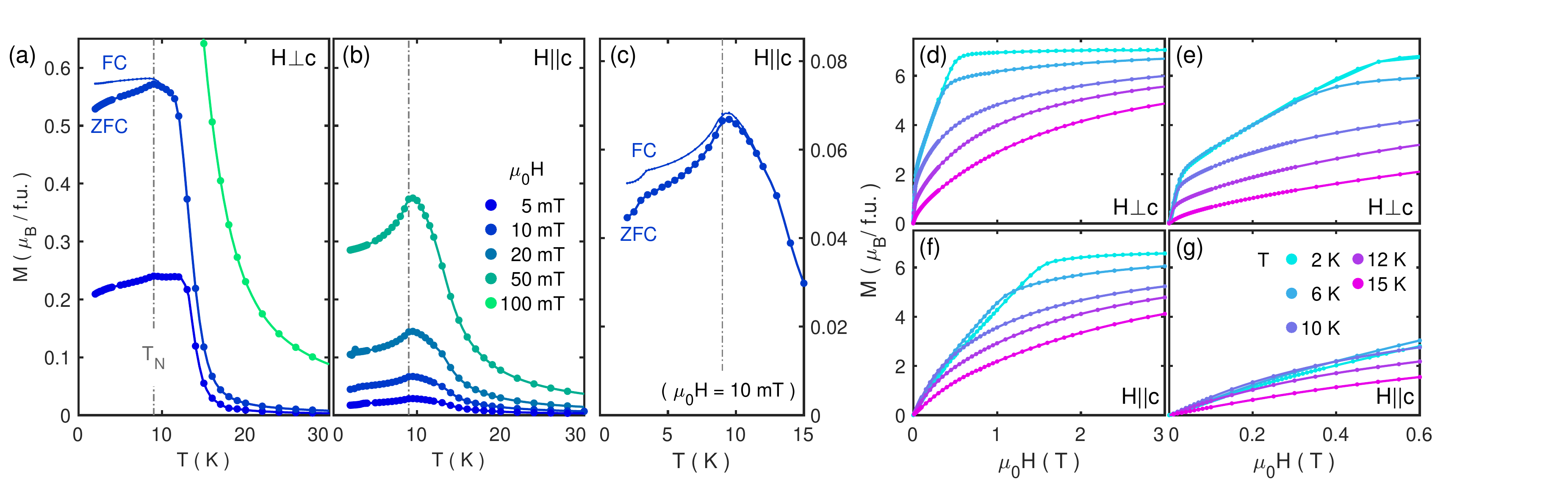}
\caption{\label{Fig2} (color online). Direction-resolved SQUID magnetization measurements of \eca. Low-temperature thermal variation of the magnetization in magnetic fields between 5 and 100\,mT applied along the (a) $a$ axis, and (b) $c$ axis. For 10\,mT data, an enlarged comparison of field-cooled (FC) and zero-field-cooled (ZFC) data is shown in (c). All data were recorded upon warming. Field sweeps of the magnetization at temperatures between 2 and 15\,K, in a field oriented (d) within the basal plane and (f) almost parallel to the $c$ axis. Panels (e) and (g) give a magnified view of the respective low-field data.}
\end{figure*}

As determined in the present study, the magnetic propagation vector in the antiferromagnetic state of \eca~is $\textbf{q}_\textrm{m}=(0,0,0.5)$, and thus the magnetic unit cell contains only two, antiparallel, magnetic moments $\bf{m}$ at fractional coordinates $\textbf{r}_1=(0,0,0)$ and $-\textbf{m}$ at $\textbf{r}_2=(0,0,1)$.  The magnetic structure factor vector in the reference frame of the crystal is therefore
\begin{equation}
\textbf{M}_\textrm{cryst}=\sum_j \textbf{m}_j \exp({\textrm{i}\textbf{q}_\textrm{m}\cdot\textbf{r}_j}) = 2\,\textbf{m}\,.
\end{equation}
The determination of the magnetic structure thus reduces to the measurement of the orientation of $\bf{m}$ through its projections onto the cross product of linear polarization vectors [see Eq.~(\ref{Eq1})].

For a continuous rotation of the angle $\eta$ of the incident linear polarization $\hat{\bf{\epsilon_i}}$, a pair of diamond phase plates (300\,$\mu$m thickness, $\approx48\,$\% transmission) is inserted in the incoming beam~\cite{Francoual2013,Strempfer2013}. The angle $\eta'$ ($\hat{\bf{\epsilon_f}}$) of linearly polarized components of the scattered beam was selected by rotation of a Cu (110) analyzer crystal (at a scattering angle of $2\theta_\mathrm{A}\approx90^\circ$) around $\hat{\bf{k}}_f$~\cite{supplemental}. The polarization state of an x-ray beam is described by its Poincar\'e-Stokes vector ($P_1$,$P_2$,$P_3$), as defined in the Supplemental Material~\cite{supplemental}. In this REXS experiment, we measured the parameters $P_1'$ and $P_2'$ of the scattered beam as a function of the angle $\eta$ of the incident linear polarization~\cite{supplemental}.

We used the Quantum Espresso suite~\cite{Giannozzi2009} to perform electronic structure calculations of EuCd$_2$As$_2$. To treat the strong electronic correlations in the highly localized rare-earth 4$f$ states, a Hubbard term $U=3.1$\,eV was included. This term adds an energy cost to the site-hopping of the 4$f$ electrons, which yields a magnetic moment that is consistent with the spin state $S = 7/2$ of localized Eu$^{2+}$. To account for the possible spin splitting of the bands due to the magnetic ions, the local spin density approximation was used as the exchange correlation. Given the presence of the heavy element Cd, relativistic pseudopotentials were also implemented in the plane wave electronic structure code to account for spin--orbit coupling. A Monkhorst--Pack \textbf{k}-point sampling mesh of $8\times8\times6$ was used~\cite{Monkhorst1976}.

\section{III. Experimental Results}
\subsection{A. Magnetization}
The temperature- and field-dependence of the magnetization is summarized in Fig.~\ref{Fig2}. The temperature sweeps reveal a magnetic ordering transition with unusual characteristics. A strong divergence of the magnetization is seen below about 20\,K, which is arrested at around 13\,K. The transition to antiferromagnetic order is signaled by a sharp change in slope at around 9\,K.

The magnetic behavior is consistent with previous studies~\cite{Artmann1996,Schellenberg2011}, and had been interpreted as two distinct, consecutive, ferro- and antiferromagnetic phases by some authors~\cite{Artmann1996}. However, these earlier studies had not been  direction-selective. In panels (a) and (b) of Fig.~\ref{Fig2}, we compare data obtained with the magnetic field parallel to the $a$ and $c$ axis, respectively. This reveals that the magnetic state is strongly anisotropic, with an in-plane susceptibility about 7 times larger than the out-of-plane susceptibility. The in-plane response is also \textit{qualitatively} different, in that it appears to saturate and form a plateau, 2--3\,K above the onset of antiferromagnetism.

\begin{figure*}
\includegraphics[width=2.02\columnwidth,trim= 4pt 7pt 29pt 0pt, clip]{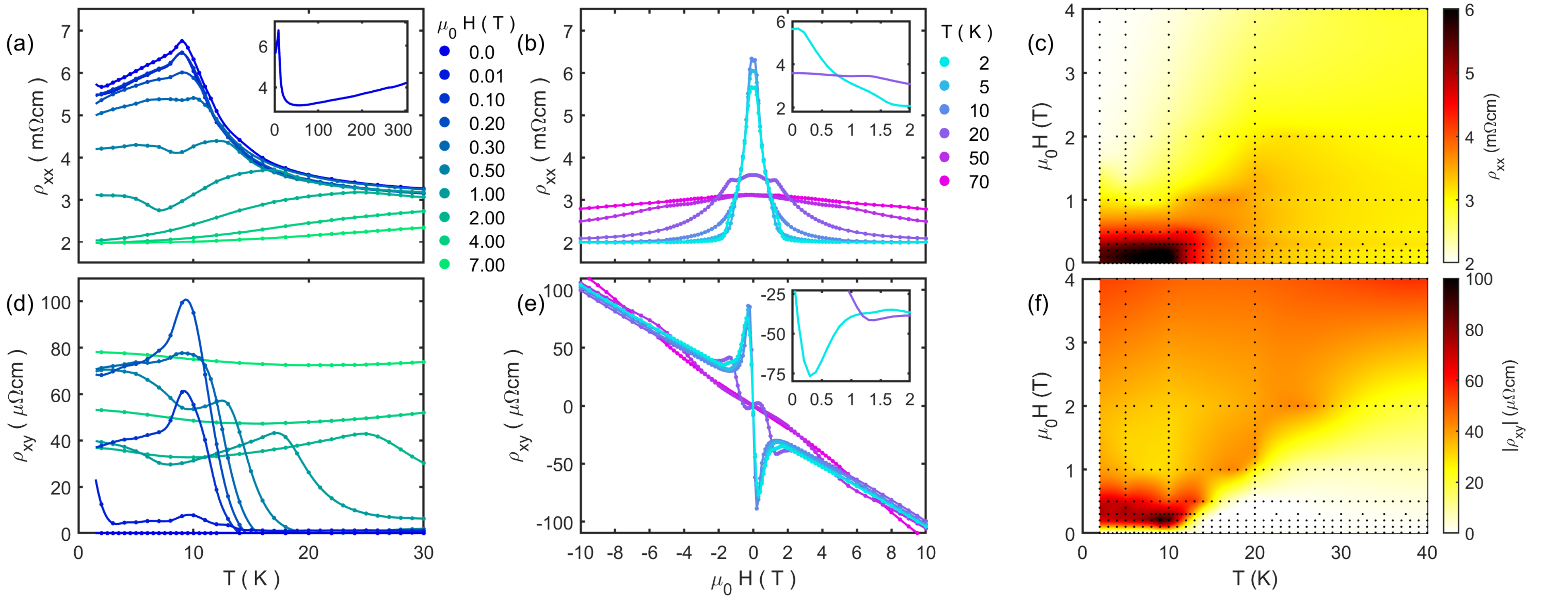} 
\caption{\label{Fig3} (color online). Alternating-current six-point electronic transport data of \eca, recorded with a 5\,mA, 30\,Hz excitation and magnetic fields applied along the $c$ axis. The characteristics of the in-plane longitudinal resistivity $\rho_{xx}$, and transversal (Hall) resistivity $\rho_{xy}$ are summarized in the top and bottom panels, respectively. (a),(d) Thermal variation at magnetic fields between 0 and 7\,T. (b),(e) Field sweeps at temperatures between 2 and 70\,K. The insets give a magnified view of subtle features in the low temperature data. (c),(f) Phase diagrams represented by color maps of interpolated data. Black markers indicate the actual location of data points.}
\end{figure*}

The comparison of temperature sweeps recorded in field-cooled (FC) and zero-field-cooled (ZFC) conditions shows a splitting, see Fig.~\ref{Fig2}\,(a) and (c), which is more significant for $\bf{H}\perp \bf{c}$ ($\approx10\%$). For this geometry, the effect is consistent with the preferred alignment of antiferromagnetic domains, as observed in our REXS study and discussed in Section IV. In-plane rotations of AFM domains would not be expected to cause a FC--ZFC splitting for $\bf{H}\parallel\bf{c}$, and the weak splitting observed in Fig. 2(c) is most likely attributable to a small misalignment of the crystal giving contamination from the $\bf{H}\perp\bf{c}$ signal. For very low applied fields, $\mu_0H<50\,$mT, another anomaly appears in the out-of-plane susceptibility, which is seen as an additional decrease of the signal [below about $3\,$K at the lowest field, 5\,mT, see Fig.~\ref{Fig2}\,(c)]. This feature has not been observed previously and its origin is presently not known.

Field sweeps of the in-plane and out-of-plane magnetization are shown in Fig.~\ref{Fig2}\,(d)--(g). As had been reported previously, the saturation magnetization approaches the value of $7\,\mu_\text{B}$ expected for an ideal Eu$^{2+}$ ($4f^7$, $S=\sfrac72$, $L=0$) state~\cite{Artmann1996,Schellenberg2011}.
At base temperature (2\,K) and with a field applied along the $c$-axis, this saturation is achieved already in a relatively low field of $1.55\,$T. The $\bf{H}\parallel \bf{c}$ curve shows no evidence for any spin-flop transitions, but there is a subtle change in slope around 200\,mT (at 2\,K). By contrast, the application of a mere $50\,$mT within the basal plane sees a jump of the magnetization to around one third of its saturation value, followed by a steep linear increase and saturation near $0.5\,$T. An interpretation of these characteristics is provided below, in light of the REXS results.

\subsection{B. Electronic transport}
In the top and bottom panels of Fig.~\ref{Fig3}, we summarize the low temperature characteristics of the longitudinal resistivity $\rho_{xx}$ and transversal (Hall) resistivity $\rho_{xy}$, respectively. Above $\sim$80\,K, the resistivity temperature sweep indicates a weak semimetallic behavior, see Fig.~\ref{Fig3}\,(a) and inset. However, at lower temperatures, the resistivity increases by up to 100\% and forms a sharp peak at $T_\text{N}=9.5\,$K. This is reminiscent of the corresponding magnetization characteristics shown in Fig.~\ref{Fig3}\,(a), although the anomalous increase in resistivity sets in at even higher temperatures $\approx50\,\text{K}\approx 5\,T_\text{N}$. When a magnetic field is applied (out-of-plane), this resistivity peak splits into two broad features: one with a leading edge that moves to higher temperatures in higher fields, and another, separated by a dip (minimum) of $\rho_{xx}$. By comparison with the magnetization data, it can be inferred that this minimum corresponds to the magnetic phase boundary, and, accordingly, disappears with the complete spin alignment at $1.55\,$T --- see Fig.~\ref{Fig2}\,(f).

The field sweeps of this resistivity signal, Fig.~\ref{Fig3}\,(b), emphasize a strong negative magnetoresistance, corresponding to a $\approx75\%$ reduction of the measured signal at 2\,K. Within the magnetically ordered phase, the negative magnetoresistance is clearly associated with the spin-alignment, as it abruptly ends at the magnetization saturation field of $1.55\,$T. Another magnetoresistive feature is observed at temperatures \textit{above} the ordering transition. It does not set in continuously, but is delineated by a clear kink in the field sweep --- see data for 20\,K in Fig.~\ref{Fig3}\,(b). 

The features discussed above can all be recognized in the phase diagram of interpolated resistivity data presented in Fig.~\ref{Fig3}\,(c). Here, the magnetically ordered phase corresponds to a dark patch with a black-to-orange gradient. The regime of strong negative magnetoresistance appears as a yellow-to-white gradient and forms a diagonal phase line from the ordered phase towards high fields and high temperatures.

The characteristics of the Hall resistivity are summarized in the corresponding lower panels (d)--(f) of Fig.~\ref{Fig3} and are governed by a similar phase diagram. In particular, the effective charge transport is hole-like and is superposed by \textit{two} anomalous contributions: one which is confined to the magnetically ordered phase, and a distinct Hall contribution associated with the spin-polarized state for $T > T_\textrm{N}$, which is clearly delineated by a diagonal phase line.

\begin{figure}
\includegraphics[width=1\columnwidth,trim= 0pt 0pt 0pt 0pt, clip]{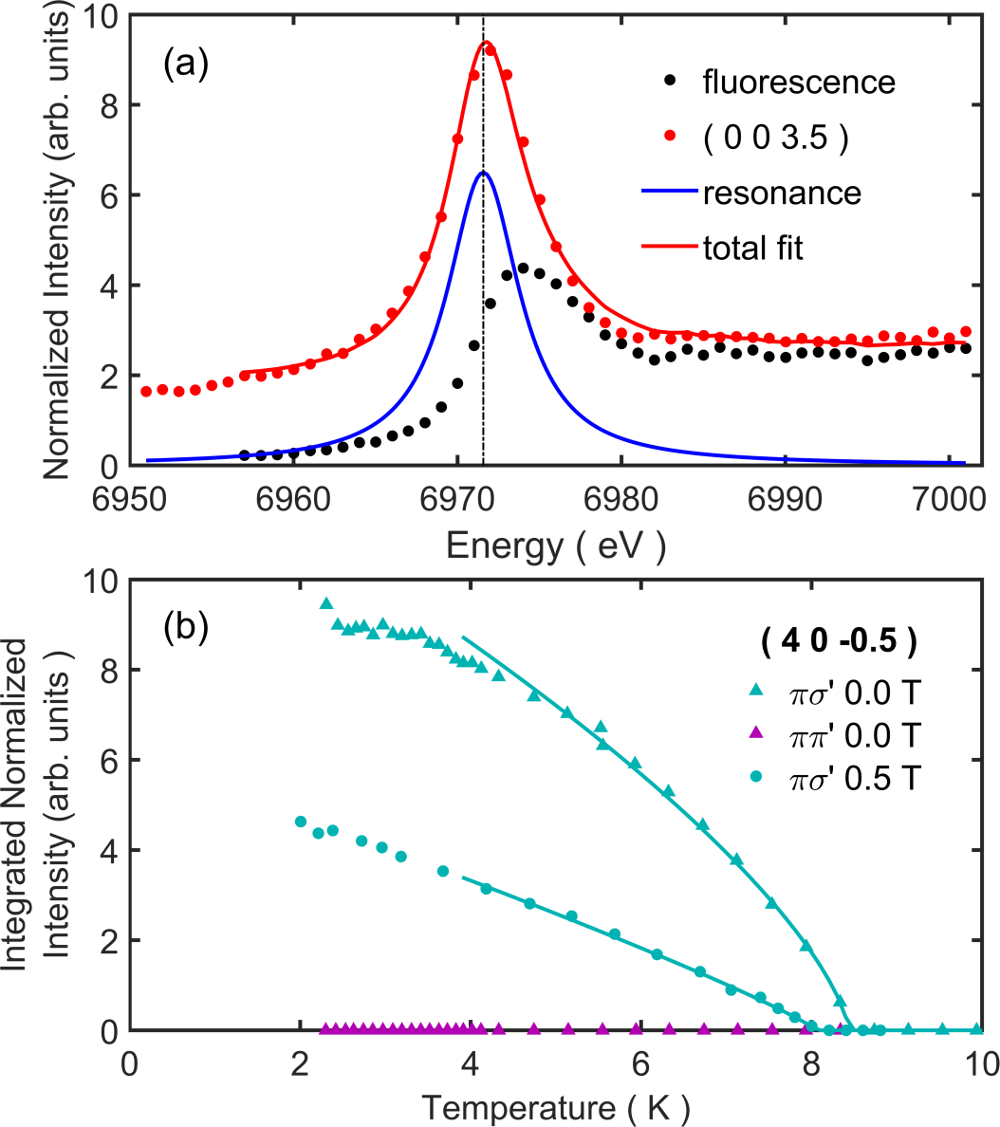}
\caption{\label{Fig4} (color online). (a) Energy scans at the Eu $L_3$ resonance, recorded without polarization analysis of the scattered beam. The magnetic resonance at the $(0,0,3.5)$ reflection can be decomposed into a contribution due to x-ray fluorescence (black markers) and a Lorentzian peak centered in the pre-edge region, at 6972\,eV. (b) Thermal variation of the $(4,0,-0.5)$ integrated peak intensity, recorded at zero field and with $0.5\,$T applied along the $c$ axis. The solid lines indicate fits of a power law $(T_\text{N}-T)^{2\beta}$ to the data ($T_\mathrm{N}=8.43(3)$, $\beta=0.34(1)$ for $\pi\sigma'$ at 0\,T). In the $\pi\pi'$ channel no magnetic peak is observed.}
\end{figure}

\begin{figure}
\includegraphics[width=1\columnwidth,trim= 0pt 0pt 0pt 0pt, clip]{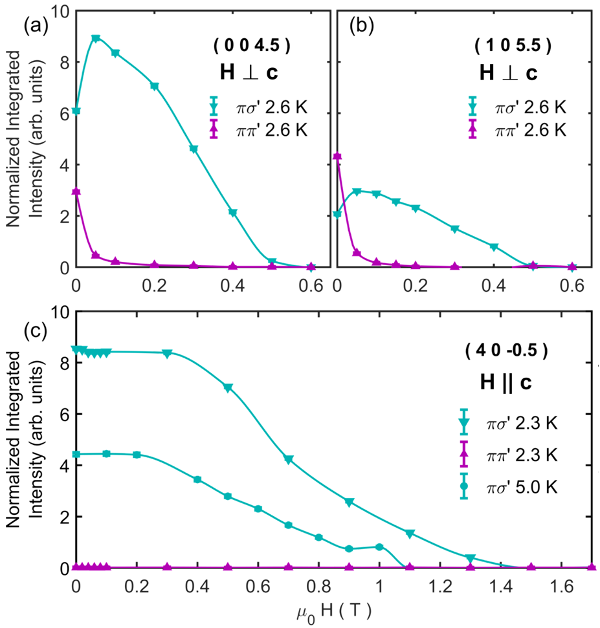}
\caption{\label{Fig5} (color online). Field dependence of \eca~resonant magnetic x-ray scattering observed in the $\pi\sigma'$ and $\pi\pi'$ polarization channels. (a)--(b) For fields applied in the $a$-$b$ plane of the crystal, the scattered intensity abruptly switches from $\pi\pi'$ to $\pi\sigma'$ character at $\approx 50\,$mT and is then fully suppressed at $\approx0.5\,$T. (c) When a magnetic field is applied parallel to the $c$ axis, $\approx 1.5\,$T is required to fully spin-polarize the sample. Solid lines are intended as visual guides.}
\end{figure}

\subsection{C. REXS}
Upon cooling below $T_\text{N}$, scans along high-symmetry directions in reciprocal space revealed that the magnetic propagation vector in \eca~is $\textbf{q}_\textrm{m}=(0,0,0.5)$. This is consistent with other magnetic materials of the same structural family for which neutron studies have been possible, such as EuMn$_2$P$_2$~[\onlinecite{Payne2002}] and EuAl$_2$Si$_2$~[\onlinecite{Schobinger1989}]. We did therefore not perform a systematic search for additional propagation vectors. In Fig.~\ref{Fig4}\,(a), constant wave vector energy scans at the $(0,0,3.5)$ reflection (measured without polarization analysis) are compared with the x-ray fluorescence measured at a scattering angle $2\theta=90^\circ$, where charge scattering vanishes. This signal, which is equivalent to an x-ray absorption (XAS) measurement, shows no indication of the Eu$^{3+}$ valence (this would be associated with a peak about 8\,eV above the Eu$^{2+}$ $L_3$ absorption maximum~\cite{Sun2010}). In addition to this fluorescence contribution, the intensity at the (0,0,3.5) position features a Lorentzian-like peak resonating in the pre-edge region at 6972\,eV. This photon energy was selected for the remainder of the experiment.

After locating the magnetic Bragg peaks, we tracked the field- and temperature dependence of integrated magnetic intensities in the $\pi\sigma'$ and $\pi\pi'$ polarization channels. Given the form of the scattering amplitude, Eq.~(\ref{Eq1}) (with reference to the $\hat{\bf{u}}_i$ reference frame~\cite{supplemental}), these channels probe magnetic Fourier components parallel and perpendicular to the scattering plane, respectively. Fig.~\ref{Fig4}\,(b) shows the thermal variation of these integrated intensities for the $(4,0,-0.5)$ peak, in zero field and for 0.5\,T applied $5^\circ$ away from the $c$ axis.

The intensity in the $\pi\sigma'$ channel follows a clear order parameter behavior, as described by a power law. The obtained N\'eel temperatures are $T_\mathrm{N}=8.43(3)$\,K (0\,T) and 8.08(2)\,K (0.5\,T). By attenuating the beam, we confirmed that the reduction by $\approx1\,$K from the expected value is due to beam heating. We note that beam heating did \textit{not} affect the FLPA scans (discussed below), since the beam is attenuated by insertion of the phase plates. The critical exponent obtained from the power-law fit is $\beta=0.34(1)$ (0\,T). This agrees with the values expected for the 3D Ising model ($\beta=0.3250$) or 3D $XY$-antiferromagnet ($\beta=0.3460$). This points to a three-dimensional character of the antiferromagnetic state in \eca~\cite{LeGuillou1977}, although $\beta$ may also be slightly altered by beam heating.

In the $\pi\pi'$ channel, no magnetic Bragg peak is observed at the $(4,0,-0.5)$ position. This indicates that there are no sizable $\bf{q}_\text{m}$ magnetic Fourier components perpendicular to the scattering plane, i.e.~parallel to  the $c$-axis.

\begin{figure*}
\includegraphics[width=2\columnwidth,trim= 0pt 0pt 0pt 0pt, clip]{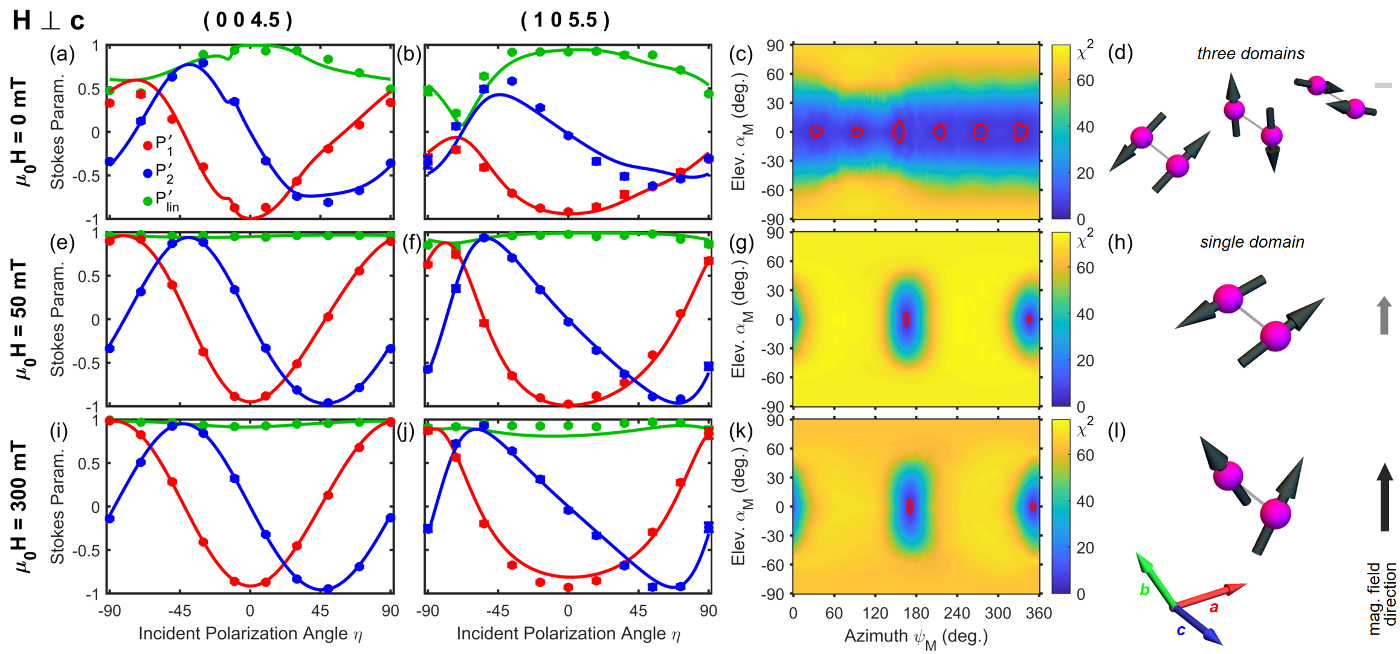}
\caption{\label{Fig6} (color online). Full polarization analysis of the $(0,0,4.5)$ and $(1,0,5.5)$ magnetic reflections of \eca, in zero field (top row) and with magnetic fields of 50 and 300\,mT applied within the $ab$ plane (middle and bottom rows). The left columns show the best fits (lines) to the Stokes parameters $P_1'$, $P_2'$ and $P_\text{lin}'=\sqrt{P_1'^2+P_2'^2}$ extracted from the polarization scans. Panels (c), (g) and (k) show the corresponding $\chi^2$ maps of least squares fits performed with the angles ($\psi_M$, $\alpha_M$) held fixed (discussion cf. main text). Panels (d), (h) and (l) illustrate the respective moment directions in adjacent layers. Where applicable, the three magnetic domains are indicated.}
\end{figure*}

\begin{figure*}[!htbp]
\includegraphics[width=2.\columnwidth,trim= 0pt 0pt 0pt 0pt, clip]{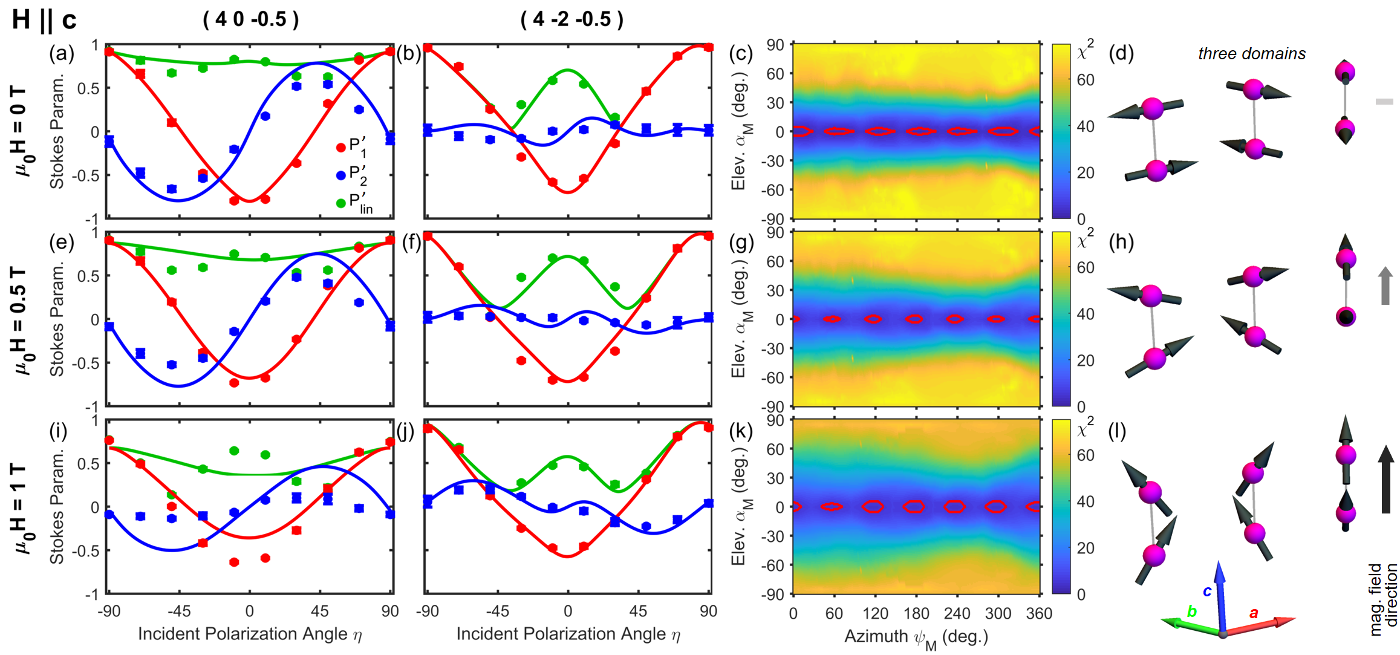}
\caption{\label{Fig7} (color online). Full polarization analysis of the $(4,0,-0.5)$ and $(4,-2,-0.5)$ magnetic reflections of \eca, in zero field and with magnetic fields of 0.5 and 1\,T applied almost parallel to the $c$ axis. The figure is organized in analogy to Fig.~\ref{Fig6}. In all cases, fits using three domains were superior to models using a single domain or an isotropic distribution of azimuthal moment orientations. In applied fields, the magnetic intensity at the $\bf{q}_\mathrm{m}=(0,0,0.5)$ magnetic Bragg peaks subsides and the relative intensity of diffuse charge scattering becomes substantial. Due to these increasing background contributions in the $\pi\pi'$ and $\sigma\sigma'$ channels, the Stokes parameter fit quality deteriorates at higher fields.}
\end{figure*}

The field dependences of three magnetic peaks at base temperature ($2.3\,$K) are shown in Fig.~\ref{Fig5}. Again, the $(0,0,4.5)$ and $(1,0,5.5)$ peaks correspond to an in-plane magnetic field, whereas for $(4,0,-0.5)$, the $c$ axis is directed $5^\circ$ away from the magnetic field axis. Both $\bf{H}\perp\bf{c}$\,-type peaks, Figs.~\ref{Fig5}\,(a) and (b), show a transition at a low field of $\approx50\,$mT, where magnetically scattered intensity changes abruptly from the $\pi\pi'$ to the $\pi\sigma'$ channel. As noted above, this implies a switching of antiferromagnetic Fourier components, which are initially perpendicular to the scattering plane, into the scattering plane. By $0.5\,$T applied in-plane, the $\pi\sigma'$ intensity has also completely disappeared. At this point the sample is fully spin-polarized, i.e.~all magnetic intensity would be located at the $\Gamma$ point of the Brillouin zone.

A different behavior is observed for the almost \,$\bf{H}\parallel\bf{c}$\,-like situation shown in Fig.~\ref{Fig5}\,(c). Here, no magnetic intensity is observed in the $\pi\pi'$ channel, i.e. the magnetic structure has no $\bf{q}_\mathrm{m}=(0,0,0.5)$ Fourier component perpendicular to the scattering plane. The intensity in the $\pi\sigma'$ channel varies little up to $\approx0.4\,$T and then decreases monotonically. By contrast to panels (a) and (b), a much larger field of $\approx1.5\,$T is required to fully spin-polarize the material (at 2.3\,K), which is consistent with the magnetization field sweeps shown above in Figs.~\ref{Fig2}\,(d)--(g).

The two left columns of Figs.~\ref{Fig6} ($\bf{H}\perp\bf{c}$) and~\ref{Fig7} ($\bf{H}\parallel\bf{c}$) show the Stokes parameters $P_1'$ and $P_2'$ and degree of linear polarization $P_\text{lin}'=\sqrt{P_1'^2+P_2'^2}$ describing the polarization state of x-rays scattered at two magnetic reflections, at zero field (top row), and with increasing strength of an applied magnetic field (second and third row). These measurements were all performed at base temperature ($\approx\,2.3\,$K). The determination of the Stokes vector ($P_1',P_2'$) from full polarization analysis is discussed in the Supplemental Material~\cite{supplemental}. Due to our observation of an anomaly at $T\approx3\,$K in the magnetic susceptibility data (see Fig. 2) we investigated specifically the possibility of a distinct low-temperature magnetic phase at zero field. However, we did not observe any variation in the polarization characteristics of the magnetic scattering (at zero field). Nevertheless, the possibility of some weak remaining beam heating to $T>3\,$K (with phase retarding plates in the beam) cannot be excluded.

As a first observation, several datasets [Figs.~\ref{Fig6}\,(a),(b) and~\ref{Fig7}\,(a), (b), (e), (f), (i), (j)] show a significant reduction in the degree of linear polarization of the scattered beam ($P_\text{lin}'$, see green markers). We identified two causes that contribute to this effect.

Firstly, the effective observed Stokes parameters may vary if the signal is contaminated by diffuse charge scattering contributions. This adds to the dipole scattering amplitude, Eq.~(\ref{Eq1}), a term of the form~\cite{Hill1996}
\begin{equation}
\hat{\bf{\epsilon}}_f\cdot\hat{\bf{\epsilon}}_i = \begin{pmatrix} 1 & 0 \\  0 & \hat{\bf{k}}_f\cdot\hat{\bf{k}}_i \end{pmatrix}  = \begin{pmatrix} 1 & 0 \\  0 & \cos2\theta \end{pmatrix}~~.
\end{equation}

The effect therefore becomes important for peaks at large scattering angles when the sample is almost spin-polarized and when antiferromagnetic scattering is weak. In Section~7 of the Supplemental Material~\cite{supplemental}, we illustrate this effect by comparison of $2\theta$-scans at the $(4,0,-0.5)$ position (at $2\theta=136.6^\circ$), in the $\pi\pi'$ and $\pi\sigma'$ polarization channels. At 1\,T, corresponding to the FLPA shown in Fig.~\ref{Fig7}\,(i), the background intensity in the $\pi\pi'$ channel is about $7\,\%$ of the magnetic signal seen in $\pi\sigma'$. In fact, when considering panels (a), (b), (e), (f), (i), (j) of Fig.~\ref{Fig7}, it can be recognized that the Stokes parameter $P_1'$ continuously becomes more ``$\cos2\eta$~-like'' and $P_2'$ becomes more  ``$-\sin2\eta$~-like'' as the magnetic field is applied. In effect, these anti-phase contributions decrease the amplitudes of both $P_1'$ and $P_2'$, and so $P_\text{lin}'=\sqrt{P_1'^2+P_2'^2}$ is no longer unity.

Nevertheless, these background contributions do not account for the decrease in $P_\text{lin}'$ in the case of Fig.~\ref{Fig7}\,(a) and (b), where the magnetic intensity is dominating the observed signal. We find that this can be modeled by taking into account the formation of magnetic domains in \eca. In general, it must be expected that a zero-field--cooled sample will contain all symmetry-equivalent magnetic domains, the scattered intensities of which add. Antiphase signals from different domains can thus introduce circular components to the scattered beam. For trigonal \eca, three equivalent magnetic domains would be expected if the magnetic moments are not aligned strictly along the $c$ axis. 

\section{IV. Analysis and interpretation}
Taking into account the above considerations, our experimental results can be explained as follows [see panels (d,h,l) of Figs.~\ref{Fig6} and Fig.~\ref{Fig7}]. The magnetic structure of \eca~consists of ferromagnetic layers which are stacked antiferromagnetically, in other words, an A-type antiferromagnet. Even though the Eu$^{2+}$ state has no orbital angular momentum, the magnetic moments experience a weak magnetocrystalline and/or dipolar anisotropy which favors spin alignment along specific directions in the $a$-$b$ plane, as already illustrated in Fig.~\ref{Fig1}. The trigonal crystal symmetry generates three equivalent antiferromagnetic domains, related by $120^\circ$ rotation (i.e.~six possible orientations of the spins), and if the domains are smaller than the volume of the sample probed by the x-rays, then the measured magnetic scattering intensity will be a sum of the intensities for each domain, weighted by the population of each domain.

For each pair of FLPA scans in Figs.~\ref{Fig6} and~\ref{Fig7}, we show a $\chi^2$ map of least squares fits with fixed magnetic moment directions [see panels (c,g,k)]. We represent the orientation of the $\bf{q}_\textrm{m}=(0,0,0.5)$ Fourier component of the magnetic structure by the azimuth and elevation ($\psi_\text{m}$, $\alpha_\text{m}$) in an orthonormal reference frame determined by the crystallographic $a$ and $c$ axes. These angles are defined and illustrated in Section~5 of the Supplemental Material~\cite{supplemental}, where we also describe the transformation of $\bf{m}_\mathrm{cryst}$ to the $\hat{\bf{u}}_i$ reference-frame. Red contours imposed onto these $\chi^2$ maps delineate the parameter range corresponding to one standard deviation from the best fit parameters.
~\\~\\
For zero applied field, both measurements [panels (a)--(b) of Figs.~\ref{Fig6} and \ref{Fig7}] yielded superior fits when the three-fold domain population was allowed to vary. The quality of the fits allows us to exclude definitively the possibility that the magnetic moments have a significant out-of-plane antiferromagnetic component. On the other hand, the fits do not reveal a unique azimuthal orientation $\psi_\text{m}$. Instead, the two crystals investigated favor different orientations, with best fits at $\psi_\text{m}\approx 37^\circ$ and $\psi_\text{m}\approx 0^\circ$ for the $\bf{H}\perp\bf{c}$ and $\bf{H}\parallel\bf{c}$ samples, respectively [see red contour lines in panels (c) of Figs.~\ref{Fig6} and \ref{Fig7}]. This may indicate a lack of significant in-plane anisotropy, or a dependence on sample history. Nevertheless, in all cases, it was \textit{not} possible to obtain an improved fit to the data with a continuous (i.e.~not three-fold discrete) distribution of azimuthal moment directions.

When a $50\,$mT in-plane magnetic field is applied, Fig.~\ref{Fig6}\,(e)--(h), $P_\text{lin}'$ is restored to unity throughout and the Stokes parameters can be perfectly reproduced by a one-domain model with the magnetic moments aligned almost perpendicular to the applied field (which happens to be close to the $a$ axis in this experiment). This behavior corresponds to a spin-flop--like redistribution of domain populations, before all moments can be continuously canted into the field direction. The small magnitude of the magnetic field which effects this \textit{domain-flop} reflects the weakness of the dipolar coupling between antiferromagnetic domains. This model also naturally explains the $\bf{H}\perp\bf{c}$ field sweeps shown in Fig.~\ref{Fig5}\,(a,b) and discussed above. Increasing the in-plane field to 300\,mT, see Fig.~\ref{Fig6}\,(i)--(l),  suppresses the antiferromagnetic Bragg peaks as the moments cant towards the field direction and intensity is redistributed to the Brillouin zone center, but the Stokes parameter scans are hardly changed.

As expected in this model, the selection of a single domain does not occur if the field does not break the three-fold symmetry. Accordingly, all results shown in Fig.~\ref{Fig7} were better modeled by imposing the three-fold domain structure. However, since this crystal was actually mounted with the $c$ axis tilted $5^\circ$ from the magnet axis, one domain (the only direction in which the magnetic moments lie both in the $a$--$b$ plane and in the scattering plane) is clearly favored by the fits. As the out-of-plane field is successively increased to 0.5 and 1\,T, the favored moment direction (close to the $a$ axis) does not change. Instead, the continuous change in Stokes parameter characteristics is attributed to the increasing relevance of the diffuse charge scattering background, as the magnetic intensity subsides.

\section{V. Discussion}
In \eca, the in-plane ferromagnetic correlations are expected to be much stronger than the inter-plane antiferromagnetic exchange interactions (six in-plane ferromagnetic nearest neighbors \textit{vs.} two more distant antiferromagnetic neighbors along the $c$ axis). This explains the continuous onset of short-range in-plane ferromagnetic correlations that starts at $T\gg T_\text{N}$, Fig.~\ref{Fig2}\,(a)--(b), and leads to an incipient ferromagnetic-like transition, before the pre-formed ferromagnetic planes lock into an antiferromagnetic stacking at $T_\textrm{N}$.

We find that the magnetic susceptibility of \eca~at $T \lesssim T_\textrm{N}$ is about 7 times larger for magnetic fields applied in the $a$--$b$ plane than for fields along the $c$ axis. For a conventional antiferromagnet, it is expected that the magnetic state is \textit{softer} for magnetic fields applied perpendicular to the moment direction. On this basis one might therefore infer an out-of-plane moment direction (as done by Wang \etal~\cite{Wang2016a}). However, in the present case, the magnetic moments are actually lying in the $a$--$b$ plane, and, owing to the three-fold domain structure, \eca~can avoid a conventional spin-flop transition irrespective of the field direction. Instead, the anisotropy of the susceptibility is mainly a measure of the easy-plane anisotropy and the fact that the spins point in different in-plane directions within each domain.   The easy-plane anisotropy is also reflected in the fact that the saturation field is smaller for $\bf{H} \perp \bf{c}$ than for $\bf{H} \parallel \bf{c}$ ($0.5\,$T \textit{vs.} $1.5\,$T).

\begin{figure}
\includegraphics[width=1\columnwidth,trim= 0pt 0pt 0pt 0pt, clip]{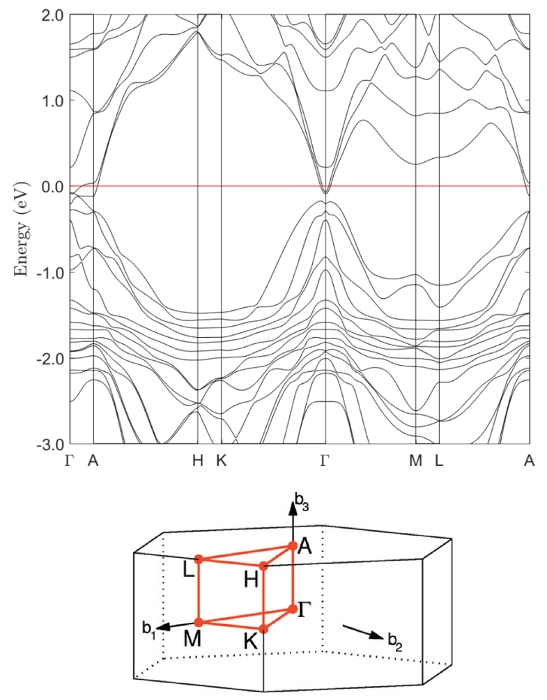}
\caption{\label{Fig8} (color online). Calculated electronic structure of \eca~along high symmetry lines of reciprocal space. The conventional labeling of high symmetry points in the Brillouin zone of the hexagonal lattice is indicated below~\cite{Setyawan2010}.}
\end{figure}

It is interesting to consider the resistivity and Hall resistivity characteristics in the light of this model. For small applied fields, anomalous Hall effect (AHE) contributions in the multi-domain state will cancel. On the other hand, at sufficiently large fields that the in-plane ferromagnetic order becomes ferromagnetically aligned in the $c$ direction, a sizable anomalous Hall effect would be expected, both above and below the (3D) magnetic ordering temperature~\cite{Nagaosa2010}. With increasing thermal fluctuations, the alignment of pre-formed 2D ferromagnetic planes will require stronger applied fields, which explains the diagonal phase line observed in Fig.~\ref{Fig3}\,(f). If the full spin alignment reduces the scattering of charge carriers from magnetic fluctuations, the above $\rho_{xy}$ argument simultaneously explains the corresponding ``inverse'' $\rho_{xx}$ phase diagram of Fig.~\ref{Fig3}\,(c).

\begin{figure*}
\includegraphics[width=2.0\columnwidth,trim= 0pt 0pt 2pt 0pt, clip]{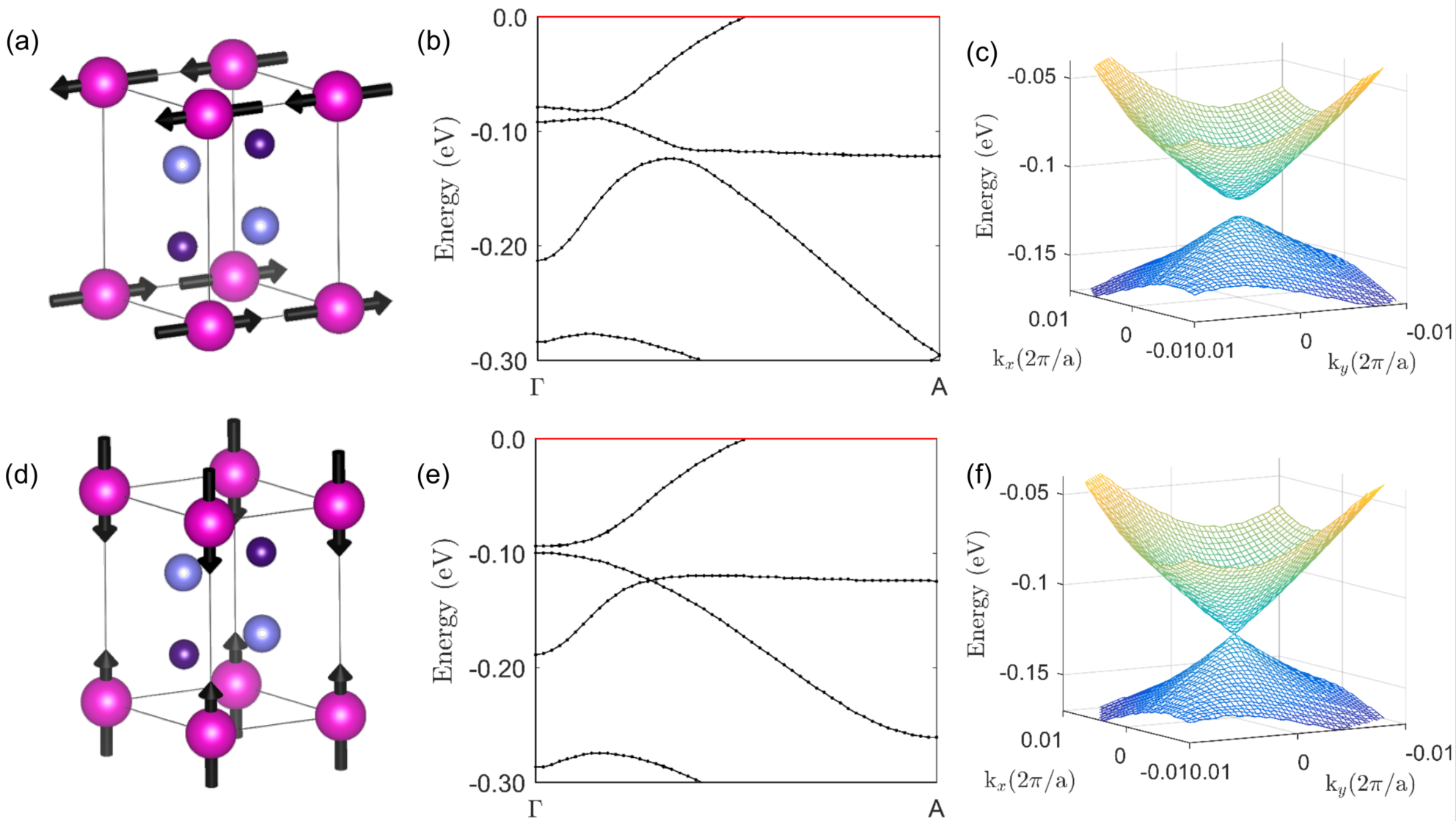}
\caption{\label{Fig9} (color online). Calculated electron bands near $E_\textrm{F}$ for A-type antiferromagnetic order in EuCd$_2$As$_2$. (a)--(c) Spins lying in the $a$--$b$ plane, as found experimentally, and (d)--(f) spins along the $c$ axis. The calculated dispersion $E(k_z)$ along the $\Gamma$--A high symmetry line corresponding to these magnetic states is presented in (b) and (e), and detailed views of $E(k_x,k_y)$ in the vicinity of the Dirac band crossing are shown in (c) and (f). This crossing is gapped for in-plane spins, as shown in (c), but is protected by $C_3$ symmetry for spins parallel to the $c$ axis resulting in Dirac points at $\textbf{k}=\pm(0,0,0.16)$, as shown in (f).}
\end{figure*}

On the other hand, this explanation does \textit{not} encompass the transport characteristics \textit{within} the antiferromagnetically ordered phase.  Contrary to observation, no large AHE contribution would be expected in the antiferromagnetic state (but only when a significant net ferromagnetic component is induced by an applied magnetic field). The insets of Figs.~\ref{Fig3}\,(b) and (e) show both large longitudinal and transversal resistivity contributions that are effectively proportional to the antiferromagnetic component measured by REXS, see Fig.~\ref{Fig5}\,(c). This implies a more subtle coupling between the magnetic state and the electronic transport.

\section{VI. \textit{Ab initio} electronic structure}

Density functional calculations of the $\textbf{q}_\textrm{m}=(0,0,0.5)$ A-type antiferromagnetic state of \eca~were performed in the magnetic unit cell (i.e.~in a cell doubled along the $c$ axis relative to the crystallographic cell). The orientation of the in-plane Eu magnetic moments was defined as shown in Fig.~\ref{Fig1}\,(a). In Fig.~\ref{Fig8} we show the calculated electronic band dispersion along high-symmetry directions in the Brillouin zone. At binding energies between 1.5--2.0\,eV, the system is dominated by weakly dispersing Eu 4$f$ bands. This suggests that these electrons are strongly localized, which is consistent with the prevailing picture for divalent Eu. The calculated ordered magnetic moment per Eu of the $4f$ states is $\mu = 6.93$\,$\mu_\textrm{B}$, corresponding to $\mu_\text{eff}=$ 7.86\,$\mu_B$, in good agreement with the estimates from magnetometry.

This binding energy of Eu 4$f$ bands is highly sensitive to the magnitude of the Hubbard-like $U$ parameter~\cite{Schlipf2013,An2011}. As typically seen in DFT calculations~\cite{Kotliar2004}, neglect of the electronic correlations $U$ would lift the 4$f$ bands to the Fermi level. Under these circumstances, one would predict \eca~to be a good metal~\cite{Larson2006,Kunes2005,Ingle2008}. Indeed, earlier computational work by Zhang \textit{et al.} suggested that this is the case in isostructural EuCd$_2$Sb$_2$, see Refs.~\onlinecite{Zhang2010} and \onlinecite{Zhang2010a}. This scenario is appealing, because it allows to attribute magnetoresistive effects to the shifting of the Eu 4$f$ bands across the Fermi level, i.e.~the magnetic bands would themselves contribute to the charge transport.

However, preliminary ARPES measurements on \eca~have ruled out this possibility~\cite{Schroeter2016}. Instead, the 4$f$ bands are observed $\approx 1.5$\,eV below the Fermi level, and we have used a Hubbard parameter $U=3.1$\,eV to reflect this in our calculation. This is also consistent with the semimetallic in-plane and inter-plane behavior of \eca~that was previously reported by Wang \textit{et al.}~\cite{Wang2016}. The orbital character analysis of our density functional study confirms that the conduction and valence bands are dominated by the Cd 5$s$ and As 4$p$ states, respectively (see Section~8 of the Supplemental Material~\cite{supplemental}). This implies that charge transport is confined to the Cd--As double-corrugated trigonal layers, which are spaced by insulating magnetic Eu layers.

Our calculations also show that the conduction and valence bands exhibit a two-fold degeneracy without spin splitting, despite the presence of the magnetic Eu species. Moreover, due to the strong spin--orbit coupling in the heavy-element Cd--As sublattice, these bands are inverted at the $\Gamma$ point in the Brillouin zone, i.e.~at this point the bands with Cd 5$s$ orbital character reside below that with As 4$p$ character. All these qualities bear striking resemblance to those of the three-dimensional topological Dirac semimetal Cd$_3$As$_2$: (1) the electronic properties are determined by the Cd 5$s$ and As 4$p$ states, (2)~there is an inversion of the aforementioned bands at the $\Gamma$ point, and (3) the bands are all doubly-degenerate~\cite{Liu2014,Neupane2014,Wang2013}.

In light of these analogies, we searched the Brillouin zone for similar topologically non-trivial features to those found in Cd$_3$As$_2$, and indeed identified a linear band crossing along the $\Gamma$--A high symmetry line at $k_z=\pm0.16\,$\AA$^{-1}$, with a gap of~$\approx10$\,meV, see Fig.~\ref{Fig9}\,(a)--(c). This energy gap is not induced by broken time-reversal symmetry because spatial inversion symmetry is simultaneously broken in \eca, and so the product of the two symmetries is conserved. Instead, this gap arises because the $C_3$ symmetry along the $\Gamma$--A line which protects the Dirac crossing point is broken by the magnetic order in which, as found here, the spins lie parallel to the $a$--$b$ plane. This gap will close if the $C_3$ symmetry is preserved, e.g.~for a magnetic structure in which the spins lie along the c-axis as shown in Fig.~\ref{Fig9}\,(d)--(f). As the spins lie in the $a$--$b$ plane this result suggests that the magnetic ground state of EuCd$_2$As$_2$ does not harbor massless Dirac fermions. The extent to which massive Dirac fermions influence the transport in EuCd$_2$As$_2$ remains to be established and, in addition, the possibility of surface states due to the non-trivial topology in the electronic structure cannot be excluded~\cite{Hua2018}.

\section{VII. Conclusions}
In conclusion, by comprehensive REXS experiments in applied magnetic fields, we have formed a consistent picture of the magnetic ground state of \eca. This demonstrates a successful application of REXS, in a material where neutron diffraction would be very difficult without isotopic enrichment with both Eu-153 and Cd-114. Using our understanding of the magnetic order we demonstrated that several distinct features in the resistivity can be understood in terms of field-induced domain alignment and/or spin reorientation.

Recently, Wang \etal~reported an electrical and optical conductivity study of \eca~in which, based on resistivity measurements, an out-of-plane magnetic moment direction was postulated~\cite{Wang2016a}. The present findings show conclusively that the spins in fact lie in the $a$--$b$ plane.

It will be of great interest to perform photoemission and quantum oscillation measurements on \eca~to corroborate our computational study of the band topology and to establish the extent to which Dirac-like fermions influence the charge transport. Our results suggest that the CaAl$_2$Si$_2$ structure could  host Dirac fermions that couple to a magnetically-ordered state at zero field. A possible route to achieve this is to search for related rare-earth based magnetic materials in which the magnetic order preserves the $C_3$ symmetry along the $\Gamma$--A line.

Note added: During the preparation of this manuscript a preprint appeared~\cite{Hua2018} in which electronic structure calculations were reported for EuCd$_2$As$_2$. The authors consider various possible magnetic structures, though not the one found in our experiments. The authors do investigate the A-type antiferromagnetic structure with moments aligned parallel to the $c$ axis, Fig.~\ref{Fig9}\,(d), and for this case the calculated band dispersion in Ref.~\onlinecite{Hua2018} is consistent with our results shown in Figs.~\ref{Fig9}\,(e)--(f).

\section{Acknowledgments}
\begin{acknowledgments}
The authors would like to thank Dr. Roger Johnson, Dr.~Niels Schr\"oter, Dr.~Fernando de Juan and Prof. Yulin Chen (Oxford) for useful discussions. We are also grateful to Dr. David Johnston (Ames Laboratory) for helpful comments and to David Reuther (DESY) for technical support. This work was supported by the U.K. Engineering and Physical Sciences Research Council (grant no. EP/J017124/1), the Chinese National Key Research and Development Program (2016YFA0300604) and the Strategic Priority Research Program (B) of the Chinese Academy of Sciences (Grant No. XDB07020100). Part of this research was carried out at PETRA-III at DESY, a member of Helmholtz Association (HGF). MCR is grateful to the Oxford University Clarendon Fund for the provision of a studentship, and to the LANL Director's Fund and the Humboldt Foundation for financial support.
\end{acknowledgments}

\bibliography{EuCd2As2bib}

\cleardoublepage
\onecolumngrid
\appendix

\begin{center}

\Large
Supplemental Material:

\vspace{0.8cm}
\large
{\bf Coupling of magnetic order and charge transport in the candidate Dirac semimetal EuCd$_2$As$_2$}

\vspace{0.5cm}
\normalsize
M. C. Rahn,$^{1,}$* J.-R. Soh,$^1$ S. Francoual,$^2$ L. S. I. Veiga,$^{2,\dagger}$ J. Strempfer,$^{2,\ddagger}$\\ J. Mardegan,$^{2,\mathsection}$ D. Y. Yan,$^3$ Y. F. Guo,$^{4,5}$ Y. G. Shi,$^3$ and A. T. Boothroyd$^{1,\mathparagraph}$
\vspace{0.2cm}
\small
 
$^1${\it Department of Physics, University of Oxford, Clarendon Laboratory, Oxford, OX1 3PU, United Kingdom}\\[1pt]
$^2${\it Deutsches Elektronen-Synchrotron (DESY), Notkestra\ss e 85, 22607 Hamburg, Germany}\\[1pt]
$^3${\it Beijing National Laboratory for Condensed Matter Physics, \\Institute of Physics, Chinese Academy of Sciences, Beijing 100190, China}\\[1pt]
$^4${\it School of Physical Science and Technology, ShanghaiTech University, Shanghai 201210, China}\\[1pt]
$^5${\it CAS Center for Excellence in Superconducting Electronics (CENSE), Shanghai 200050, China}\\[1pt]

\end{center}
\vspace{1cm}

\twocolumngrid

\begin{center}
{\bf 1. Laboratory single crystal x-ray diffraction}
\end{center}

As described in the manuscript, the flux-grown single crystals were characterized by laboratory single-crystal x-ray diffraction to confirm the structure and stoichiometry. The measurement was performed on a Agilent Supernova kappa-diffractometer, using a Mo $K_\alpha$ micro-focused source and an Atlas S2 CCD area detector \cite{Agilent}.

The trigonal crystal structure (space group $P\bar{3}m1$) of \eca~had been originally determined by Artman \etal~\cite{Artmann1996}, and is shown in Fig.~S1\,(c,d). The present single crystal diffraction dataset comprised 1077 Bragg reflections, which were indexed in a trigonal cell [$a,a,c,90^\circ,90^\circ,120^\circ$] with $a\approx4.44\,$\AA~and $c\approx7.33\,$\AA. Intensity integrated over a margin perpendicular to high symmetry planes of reciprocal space is shown in Fig.~S1\,(a). In all directions, the average mosaicity was smaller than the instrumental resolution ($\approx0.6$--$0.9^\circ$), which indicates the high crystalline quality of these samples. Rocking scans on the Huber diffractometer at beamline P09 confirmed a mosaicity of 0.15$^\circ$ or better.

We also performed a refinement of the integrated Bragg intensities (FullProf algorithm~\cite{Rodriguez1993}). The best fit to the data is shown in Fig.~S1\,(b). Due to the heavy elements, absorption effects are strong and the comparison factor of equivalent reflections is high, $R_\text{int}=14\%$. Nevertheless, a satisfactory refinement ($R_\text{F}=3.74$) was achieved. The obtained atomic parameters, listed in Table~I., are consistent with literature~\cite{Artmann1996,Schellenberg2011}. The refinement of atomic occupation factors confirmed the ideal stoichiometry of the sample.

\begin{table}
\caption{\label{TabS1} Results of the refinement of integrated single crystal x-ray diffraction intensities of \eca~(as illustrated in Fig.~S1): Wyckoff sites of space group $P\bar{3}m1$, atomic positions, thermal parameters and site occupations. The observed peaks were indexed in a trigonal cell with lattice parameters $a\approx4.44\,$\AA~and $c\approx7.33\,$\AA.}
\begin{ruledtabular}
\begin{center}
\begin{tabular}{ l r c c c c c }
   & Wyck. & $x$ & $y$ & $z$ & $B$ & occ. (\%)\\
\hline
Eu & 1a & 0 & 0 & 0 & 1.08(6) & 100 \\
Cd & 2d & $\sfrac13$ & \sfrac23 & 0.6333(3) & 1.27(7) & 100 \\
As & 2d & $\sfrac13$ & \sfrac23 & 0.2463(5) & 1.09(7) & 100 \\
\end{tabular}
\end{center}
\end{ruledtabular}
\end{table}

\begin{figure*}
\includegraphics[width=1.9\columnwidth,trim= 0pt 0pt 0pt 0pt, clip]{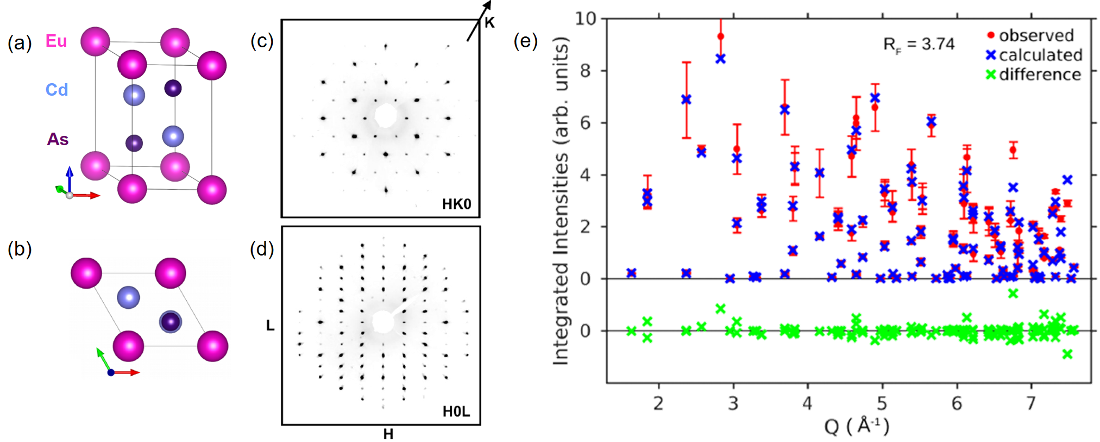}
\justify{\label{FigS1} Fig.~S1 (color online). Room temperature Mo $K_\alpha$ four circle x-ray diffraction of a $\approx\,200\,\mu\text{m}$ \eca~crystallite. (a,b) Perspective and top views of the trigonal structure of \eca. (c) Integration of the dataset over a small margin perpendicular to the ($HK0$) and, (d), ($H0L$) reciprocal lattice planes. (e) Integrated intensities sorted by momentum transfer $Q$, along with the refined values (FullProf \cite{Rodriguez1993}) and the corresponding differences. }
\end{figure*}
~\\
\begin{center}
{\bf 2. Six-point transport measurements}
\end{center}

The resistivity and Hall effect measurements reported in the main text were performed
using the alternating current transport (ACT) option of a physical properties measurement system (Quantum Design [\onlinecite{QuantumDesign}]). For all measurements, an excitation current of $5\,$mA, $30\,$Hz was applied in the $a$--$b$ plane and a magnetic field up to $10\,$T was applied along the crystallographic $c$ axis. The plate-like sample (thickness $\approx 200\,\mu m$) was contacted in a six-point geometry, such that longitudinal and transversal (Hall) voltage can be measured simultaneusly. This is illustrated in Fig.\,S2 along with the corresponding wiring of the sample holder.

\begin{figure}
\includegraphics[width=0.6\columnwidth,trim= 0pt 0pt 0pt 0pt, clip]{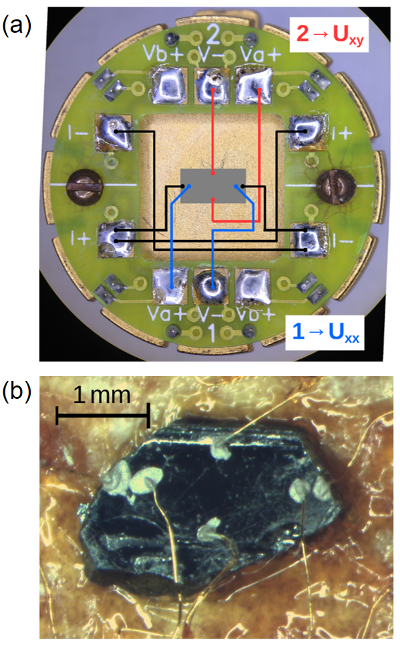}
\justify{\label{FigS2} Fig.~S2 (color online). Sample geometry for alternating current transport measurements using (Quantum Design PPMS) \cite{QuantumDesign}. (a) Wiring scheme of the PPMS sample holder using both lock-in amplifiers of the system to measure both longitudinal and transversal voltages. (b) \eca~single crystal contacted for a six-point measurement.}
\end{figure}

\begin{center}
{\bf 3. REXS setup at P09 (PETRA-III)}
\end{center}

The second experimental hutch (EH2) of the hard x-ray diffraction experimental station P09 at PETRA-III (DESY) has three unique advantages for the present study: (1) Up to high energies (3.2--14\,keV), the angle of the incident linear x-ray polarization can be rotated using a pair of diamond phase plates~\cite{Francoual2013}. (2) The non-magnetic heavy-load diffractometer holds a vertical 14\,T cryomagnet. (3) In the sample space of the variable temperature insert, there is a continuous flow of exchange gas from the liquid helium reservoir. By comparison to closed cycle refrigerators, this reduces the effects of beam heating and allows cooling of the sample down to $T_\text{base}\approx2.3\,$K.

\begin{figure*}
\includegraphics[width=2\columnwidth,trim= 0pt 0pt 0pt 0pt, clip]{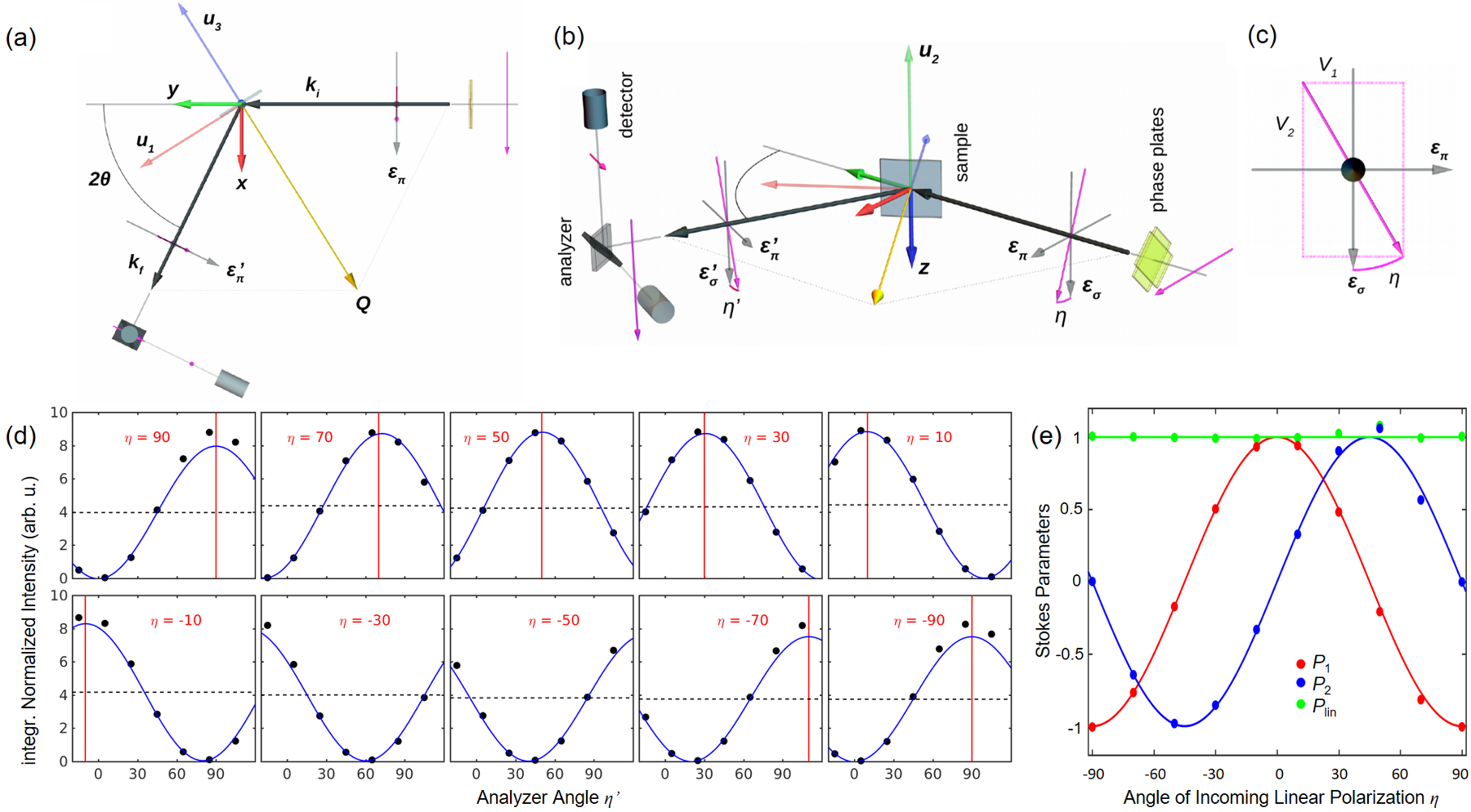}
\justify{\label{FigS3} Fig.~S3 (color online). (a) Schematic of the horizontal scattering setup in the second experimental hutch of instrument P09, PETRA-III. The scattering triangle ($\bf{k}_i$, $\bf{k}_f$, $\bf{Q}$) for an arbitrary scattering angle $2\theta$ is shown, along with the definitions of the laboratory reference frame $(\bf{x},\bf{y},\bf{z})$ and the conventional scattering reference frame $(\hat{\bf{u}}_1,\hat{\bf{u}}_2,\hat{\bf{u}}_3)$. (b) Perspective view of (a), which illustrates the orientation of the incident ($\hat{\epsilon}_\sigma$, $\hat{\epsilon}_\pi$) and scattered ($\hat{\epsilon}'_\sigma$, $\hat{\epsilon}'_\pi$) polarization vectors as well as the corresponding polarization angles $\eta$ and $\eta'$. (c) Definition of the amplitudes of the linearly polarized light, parallel ($V_1$) and perpendicular ($V_2$) to the scattering plane (view towards the beam). (d,e) Full polarization analysis of the direct beam at a photon energy of 6.972\,keV. The Stokes parameters shown in panel (e) confirm that the beam is fully linearly polarized for any setting of the phase retarder. Each set of ($P_1$, $P_2$) is obtained by fitting Eq.~\ref{EqS6} to the integrated intensities (black markers) shown in panels (d). The best fit is indicated by a blue line.}
\end{figure*}

The experimental setup of P09-EH2 is illustrated in Fig.~S3. The experiment was performed in the horizontal scattering geometry. The hard x-ray beam penetrates the cryostat through thin beryllium windows, which imposes constraints on the detector angles $2\theta$ (azimuth) and $\gamma$ (elevation). To avoid a tilting of the heavy cryostat ($\chi$ axis), the instrument was effectively used as a two-circle diffractometer (with scattering angle $2\theta$ and a parallel sample rotation axis $\omega$). Small misalignments of the sample can be compensated by an elevation of the detector out of the horizontal plane ($-5^\circ <\gamma< 5^\circ$).  In principle, this tilts the scattering plane away from that shown in Fig.~S3\,(a,b), which complicates the polarization analysis. However, as only peaks with $\gamma<1.5^\circ$ were investigated, we neglect this effect in the data analysis.

In order to investigate the magnetic state with a magnetic field applied within the basal plane ($\bf{H}\parallel\bf{c}$), a crystal was polished with a surface parallel to the [001] plane and mounted with this (001) direction in the (horizontal) scattering plane. In this setting, the $(0,0,4.5)$ and $(1,0,5.5)$ magnetic reflections were accessible. Fig.~S4 illustrates the scattering geometry for the case of $\bf{H}\parallel\bf{c}$. Here, it was desirable to access a $(h,0,-0.5)$-type magnetic Bragg peak with largest possible index $h$. To this end, a crystal was polished with a surface parallel to the $[100]$ crystal planes. A brass (non-magnetic) sample holder was then micro-machined with a $5^\circ$ tilt, such that the $(4,0,-0.5)$ direction is in the scattering plane and the field is applied only $5^\circ$ from the $c$ axis, see Fig.~S4\,(a--c). As shown in Fig.~S4\,(d), this left a narrow margin in reciprocal space in which $(h,k,-0.5)$-type peaks were accessible, delineated by the constraints in $\gamma$ and $2\theta$.

\begin{figure*}
\includegraphics[width=1.8\columnwidth,trim= 0pt 0pt 0pt 0pt, clip]{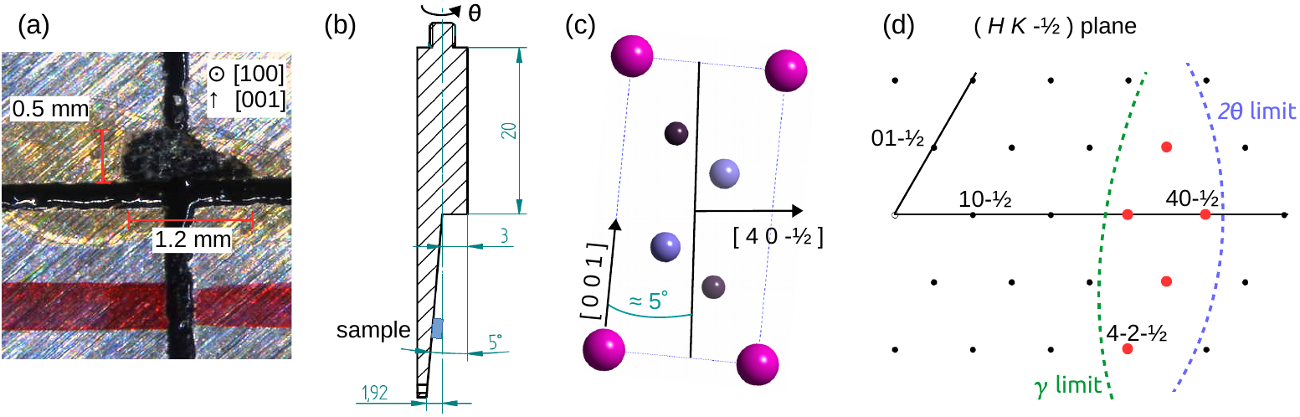}
\justify{\label{FigS4} Fig.~S4 (color online). (a) Micrograph of the crystal that was aligned and polished for $\bf{H}\parallel\bf{c}$-measurements. A surface parallel to the $[1,0,0]$ crystallographic planes has been prepared. (b) Technical drawing of a micro-machined sample holder for horizontal scattering with a 5$^{\circ}$~offset from the sample surface. (c) Unit cell of \eca~viewed along the (100) direction. The $[4,0,-0.5]$ crystallographic plane is indicated by a black line. (d) View of the $(h,k,-0.5)$ plane of reciprocal space investigated in this sample. The dashed lines mark the constraints imposed by the maximum scattering angle ($2\theta$) and the vertical aperture of the Be windows ($\gamma$). Accessible  $\bf{q}_m=(0,0,0.5)$ magnetic peaks are indicated by red markers.}
\end{figure*}
~\\
\begin{center}
{\bf 4. FLPA formalism}
\end{center}

A convenient formalism for the description of polarized x-ray scattering phenomena has been described by Detlefs \etal~\cite{Detlefs2012}. Useful discussions of full linear polarization analysis (FLPA) are also found in several reports of successful studies~\cite{Mazzoli2007,Johnson2008,Hatton2009,Shukla2012,Francoual2015}. Here we provide a brief summary of these concepts since they are relevant to the data presented in the main text.

The electric field $\bf{E}$ of an x-ray beam of energy $\hbar\omega$ and wave vector $\bf{k}$ can be decomposed into two orthogonal amplitudes $V_1$ and $V_2$.
\begin{equation}
\bf{E}(t,\bf{r})=\mathfrak{R}\left[(V_1\,\hat{\bf{\epsilon}}_\sigma+V_2\,\hat{\bf{\epsilon}}_\pi)\cdot\,e^{-i(\omega t-\bf{k}\cdot\bf{r})}\right]~~,
\tag{S1}
\end{equation}
where $\mathfrak{R}$ signifies the real part of this field. This is not relevant in the linearly polarized case, where the \textit{Jones} polarization vector,
\begin{equation}\label{App:PA:Eq01}
\hat{\bf{\epsilon}}=\begin{pmatrix}V_1\\V_2\end{pmatrix}=\begin{pmatrix}\cos\eta\\ \sin\eta\end{pmatrix}
\tag{S2}
\end{equation}
 has no imaginary components. As shown in Fig.~S3\,(a,b), the unit vectors $\hat{\bf{\epsilon}}_\sigma$, $\hat{\bf{\epsilon}}_\pi$ and $\hat{\bf{k}}_i$ form a right-handed coordinate system. For linearly polarized light, the Jones vector can thus be reduced to the polarization angle $\eta$, as indicated in Fig.~S3\,(c).

 The polarization state of the beam is described by the Poincar\'{e}-Stokes vector $(P_1,P_2,P_3)$, which is defined as
\begin{align}\label{App:PA:Eq02}
\begin{split}
P_1 &=\frac{|V_1|^2-|V_2|^2}{|V_1|^2+|V_2|^2} \\
P_2 &=\frac{|V_1+V_2|^2-|V_1-V_2|^2}{2\,(|V_1|^2+|V_2|^2)} \\
P_3 &=\frac{(|V_1-i\,V_2|^2-|V_1+i\,V_2|^2)}{2\,(|V_1|^2+|V_2|^2)}
\end{split}
\tag{S3}
\end{align}
For fully linearly polarized light, $P_3$ vanishes and the degree of linear polarization is unity,
\begin{equation}
P_\mathrm{lin}=\sqrt{P_1^2+P_2^2}\rightarrow 1
\tag{S4}
\end{equation}
By combining Eqs.~\ref{App:PA:Eq01} and~\ref{App:PA:Eq02}, it also follows that in this case
\begin{equation}\label{App:PA:Eq03}
P_1=\cos2\eta ~~~~~\text{and}~~~~~P_2=\sin2\eta ~~.
\tag{S5}
\end{equation}
These characteristics are evident in the FLPA scan of the direct beam at beamline P09 (DESY) shown in Figs.~S3\,(d,e). To obtain the Stokes parameters, the phase plates were rotated in ten steps, corresponding to a variation of the angle of incident linear polarization $\eta$ in steps of 20$^\circ$ between $-90^\circ$ and $+90^\circ$. For each incident polarization, rocking scans of the analyzer crystal were measured at seven analyzer angles $\eta'$ between $-15^\circ$ and $+105^\circ$. At the Eu $L_3$ resonance, the corresponding scattering angle of the Cu (110) reflection is $2\theta=89.4^{\circ}$. Only $\cos^2 2\theta=0.01\%$ of the components of the scattered light that are linearly polarized in the polarization analyzer scattering plane are transmitted. Such analyzer \textit{leakage} was therefore neglected in the present data analysis. The Stokes parameters $P_1$, $P_2$ can then be extracted by fitting the relation
\begin{equation}\label{EqS6}
I(\eta')=I_0+I_0\,(P_1\,\cos2\eta'\,+\,P_2\sin 2 \eta')~~.
\tag{S6}
\end{equation}

The effect of the scattering process on the polarization state can be expressed in the \textit{coherency matrix} formalism described by Detlefs \textit{et al.}~\cite{Detlefs2012}. The Stokes parameters $P_1'$ and $P_2'$ characterizing the polarization state of the scattered light can thus be simulated for a given magnetic structure.

As described in the manuscript, we also considered situations in which the probed sample volume contains several magnetic domains. To determine the effect of these contributions, Stokes parameters $P_1'^i$ and $P_2'^i$ were calculated for $N$ domains ($i=1$...$N$). Each domain contributes an intensity $I^i(\eta')\propto I_0^i$ as in Eq.~\ref{EqS6}, which add to the total signal:
\begin{equation}
I(\eta')=I_\text{tot}+I_\text{tot}\,\left( {P_1'}^\text{eff}\,\cos 2\eta'+ {P_2'}^\text{eff}\,\sin 2\eta'\right)~~,
\tag{S7}
\end{equation}
with $I_\text{tot}=\sum I_0^i$~. The \textit{effective} Stokes parameters are therefore obtained as weighted sums:
\begin{equation}
{P_1'}^\text{eff}=\frac{\sum I_0^i\,{P_1'}^i}{I_\text{tot}}~~~\text{and~~~}{P_2'}^\text{eff}=\frac{\sum I_0^i\,{P_2'}^i}{I_\text{tot}}~~~.
\tag{S8}
\end{equation}

\begin{figure}
\includegraphics[width=0.5\columnwidth,trim= 0pt 0pt 0pt 0pt, clip]{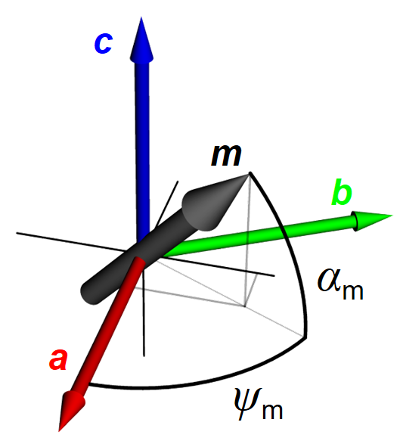}
\justify{\label{FigS5} Fig.~S5 (color online). Illustration of the azimuth $\psi_\text{m}$ and elevation $\alpha_\text{m}$ of the magnetic moment $\bf{m}$ at the origi of the unit cell, as defined in Eq.~13.}
\end{figure}

\begin{center}
{\bf 5. Reference frame}
\end{center}
As described in the main text and illustrated in Fig.~S5, the orientation of the magnetic moment at the origin of the unit cell was defined in terms of azimuthal and elevation angles in an orthogonal reference frame including the $a$ and $c$ axes of the crystal lattice:
\begin{align}
\begin{split}
\textbf{m}_\text{cryst}&=\sum_j\,\textbf{m}_j\exp({\textrm{i}\,\textbf{q}_\textrm{m}\cdot\textbf{r}_j})\rightarrow 2\,\textbf{m}\,,\\
\hat{\textbf{m}}_\textrm{cryst}&=\begin{pmatrix} \cos\alpha_\textrm{m}\,\cos\psi_\textrm{m}\\ \cos\alpha_\textrm{m}\,\sin\psi_\textrm{m}\\\sin\alpha_\textrm{m}\end{pmatrix}
\end{split}
\tag{S9}
\end{align}

In order to calculate the REXS scattering amplitude defined in Eq.~1 of the manuscript, the structure factor has to be expressed in the $\hat{\textbf{u}}_i$ reference-frame introduced by Blume and Gibbs\cite{Blume1988}:
\begin{align}\label{EuCd2As2:SEq:01}
\begin{split}
\hat{\bf{u}}_1 &= ( \hat{\bf{k}}_i+\hat{\bf{k}}_f ) / (2\,\cos\theta)~~, \\
\hat{\bf{u}}_2 &= ( \hat{\bf{k}}_i\times\hat{\bf{k}}_f ) / \sin2\theta~~,\\
\hat{\bf{u}}_3 &= ( \hat{\bf{k}}_i-\hat{\bf{k}}_f ) / (2\,\sin\theta) ~~.\\
\end{split}
\tag{S10}
\end{align}

$\hat{\textbf{m}}_\textrm{cryst}$ is transformed to the \mbox{diffractometer-,} laboratory-, and finally to the $\hat{\textbf{u}}_i$ reference-frame as follows~\cite{Busing1967}:
\begin{equation*}
\hat{\textbf{m}}_{\hat{\mathbf{u}}}=\mathrm{A}\cdot \hat{\textbf{m}}_\text{lab} = \mathrm{A}\cdot\Omega\cdot \hat{\textbf{m}}_\text{diff}=  \mathrm{A}\cdot\Omega\cdot \mathrm{U}\cdot \hat{\textbf{m}}_\textrm{cryst}
\end{equation*}
Here, $\mathrm{U}$ is the unitary orientation matrix, which describes the orientation of the crystal relative to the sample holder, as defined by Busing and Levy~\cite{Busing1967}. $A$ and $\Omega$ are rotation matrices:
\begin{equation*}
\mathrm{A}=\begin{pmatrix} \sin\theta & \cos\theta & 0 \\ 0 & 0 & -1 \\ -\cos\theta & \sin\theta & 0\end{pmatrix}~~~~\textrm{and}~~~~\Omega=\begin{pmatrix} \cos\omega & \sin\omega & 0 \\ -\sin\omega & \cos\omega & 0 \\ 0 & 0 & 1\end{pmatrix}
\end{equation*}
The transformation $\mathrm{A}$ is specific to the definition of the laboratory reference frame $(\hat{\bf{x}},\hat{\bf{y}},\hat{\bf{z}})$ shown in Fig.~S3\,(a,b). In terms of $\hat{\textbf{m}}_{\hat{\mathbf{u}}}$, the dipole REXS scattering amplitude is simply
\begin{equation}
F_{E1}\propto \left(\hat{\bf{\epsilon}}_f\times\hat{\bf{\epsilon}}_i\right)\cdot\hat{\textbf{m}}_{\hat{\mathbf{u}}}
\tag{S11}
\end{equation}
~\\

\begin{center}
{\bf 6. FLPA data analysis}
\end{center}
For the full polarization analysis scans shown in Figs.~6 and 7 of the main text, the diffractometer was first aligned on the selected magnetic peak. The scattered intensity was then investigated for a series of angles of incident linear polarization $\eta$, and for each $\eta$ at a series of angles of scattered polarization $\eta'$. For each setting ($\eta$, $\eta'$), a rocking scan of the (110) reflection of the Cu analyzer crystal was recorded. These scans were fitted by a Gaussian line shape and the integrated intensities $I(\eta')$ were stored. The Stokes parameters ($P_1'$, $P_2'$) were then extracted from this data by fitting Eq.~\ref{EqS6}.
\begin{figure}
\includegraphics[width=1\columnwidth,trim= 0pt 0pt 0pt 0pt, clip]{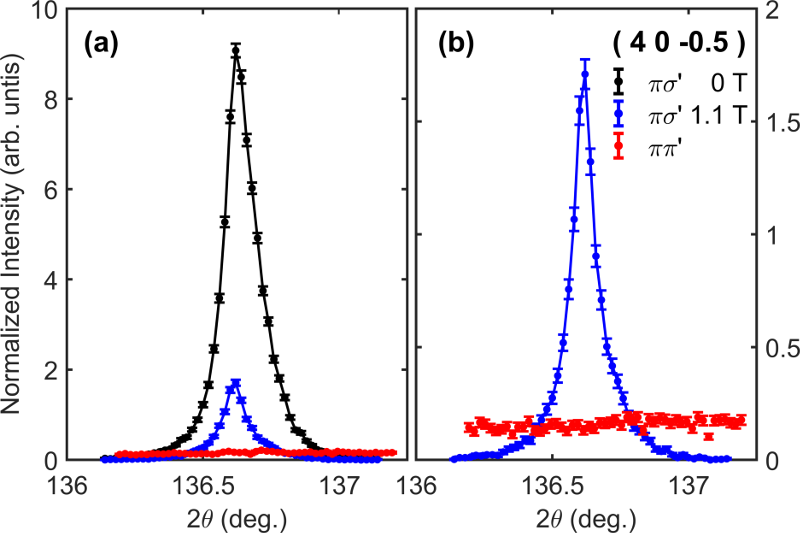}
\justify{\label{FigS6} Fig.~S6 (color online). (a,b) $2\theta$-scans at the (4,0,-0.5) position, in the $\pi\sigma'$ and $\pi\pi'$ polarization channels, at 2.3\,K, shown for zero field and at 1.1\,T. The $\pi\pi'$ channel does not feature magnetic scattering, but a field- and temperature-independent diffuse background signal. Panel (b) provides an enlarged view of the 1.1\,T data, where the relative intensity of the relative intensity of the $\pi\pi'$ channel is significant.}
\end{figure}

\begin{figure*}
\includegraphics[width=2.05\columnwidth,trim= 0pt 0pt 0pt 0pt, clip]{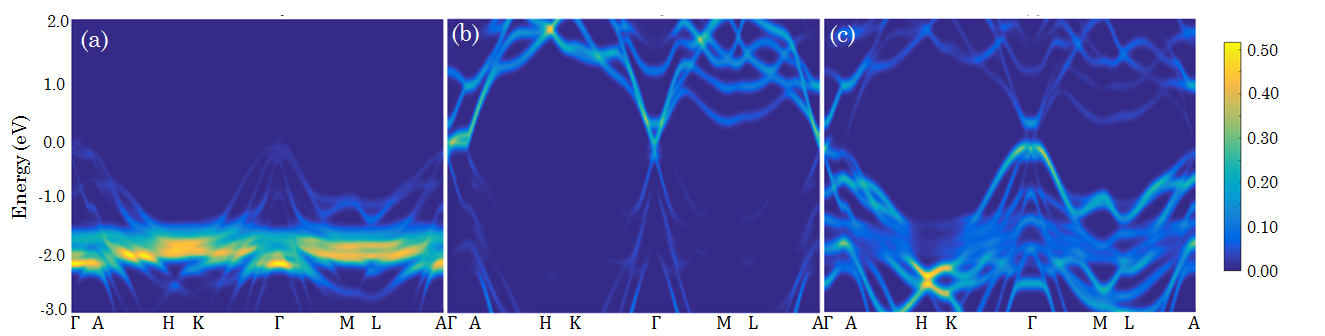}
\justify{\label{FigS7} Fig.~S7 (color online). Color plots indicating the contribution of (a) Eu $^2$F$_{5/2}$, (b) Cd $^2$S$_{1/2}$ and (c) As $^2$P$_{3/2}$ orbital characters to the band structure of \eca~shown in Fig.~8 of the manuscript.}
\end{figure*}

To determine the magnetic structure, the integrated intensities of the polarization-analyzer rocking scans were fitted directly (i.e. before extracting ${P_1'}^\mathrm{eff}$and ${P_2'}^\mathrm{eff}$), by calculating the scattering amplitudes for each combination of $\eta$ and $\eta'$. This simplified the inclusion of non-magnetic scattering contributions. In each case, all data, obtained at two independent magnetic peaks was fitted simultaneously.

The fitting parameters were (1) the azimuth $\psi_\text{M}$ and elevation $\alpha_\text{M}$ of the magnetic moment at the origin of the unit cell, (2) scale factor(s) for either one or three domains and (3) a scale factor for the amplitude of the diffuse background. Depending which configuration yielded a smaller least squares parameter $\chi^2$, the scattering amplitude was either simulated for a single moment direction, or for three magnetic domains (spaced by 120$^\circ$).

The quality of the least squares fit did not significantly deteriorate ($\chi^2\approx2$ vs. $\chi^2\approx6$) when constraining the domain population to be identical for both peaks. In Figs.~6 and 7 of the main text we show the best fits, where the domain population were \textit{not} constrained. This is justified, because the peaks are observed at different scattering angles, and thus with a different cross section of the beam on the sample. Moreover, to correct for the true center of rotation, the samples are re-centered after a move in $\omega$. Due to this translation and change of incident angle (beam cross section ca. $40\,\times\,200\,\mu m$), the scattering sample volume would therefore vary between reflections. In any case, cross-checks confirmed that the $\chi^2$ maps shown in Figs.~6 and 7 of the main text did not change \textit{qualitatively}, whether the relative domain population was constrained or not (i.e. the inferred ideal moment direction is a robust result). Constraining the population of the three domains to be \textit{isotropic} did significantly worsen the fit ($\chi^2\approx54$).

\begin{center}
{\bf 7. Diffuse charge scattering}
\end{center}
In increasing applied magnetic fields, the magnetic structure becomes spin-polarized and the magnetic REXS intensities at $\bf{q}_\mathrm{m}=(0,0,0.5)$-type Bragg peaks decrease. By contrast, the (non-magnetic) diffuse background signal is temperature- and field-independent and proportional to $\cos^22\theta$. Its relative intensity thus becomes significant, particularly for peaks at large scattering angles. For the case of the (4,0,-0.5) peak, at $2\theta=136.6^\circ$, this significantly deteriorates the FLPA fits, as is evident from Fig.~7\,(i) of the manuscript.

In Fig.~S6 we illustrate this effect with raw data of corresponding $2\theta$-scans. At zero field, the $\pi\pi'$ contribution is negligible --- hence the satisfactory FPLA fit in Fig.~7\,(a) of the manuscript. At 1.1\,T, the relative $\pi\pi'$ intensity at the peak position is 8.7\,\%, see Fig.~S6\,(b).

\begin{center}
{\bf 8. Orbital character analysis}
\end{center}

The orbital character analysis of the electronic bands obtained from the density functional calculations reveals that the $4f$ electrons which are responsible for magnetism in \eca~do not contribute to charge transport. This is illustrated by color plots of the calculated orbital contributions in Fig.~S5. Instead, the density of states at the Fermi surface is dominated by the Cd $5s$ and As $4p$ electronic states.

\bibliography{EuCd2As2Bib}

\end{document}